\documentclass{aa}
\usepackage{graphicx}
\usepackage{lipsum}
\usepackage{natbib}
\usepackage{multicol}
\usepackage{stfloats}
\usepackage{color}

\usepackage[flushleft]{threeparttable}
\usepackage{yfonts}
\usepackage{ulem}
\usepackage[dvipsnames]{xcolor}

\usepackage{txfonts}
\begin{document} 

    \title{EI Eridani: a star under the influence\thanks{Original spectroscopic data acquisition was coordinated by Albert Washuettl (former researcher at AIP) and was carried out within the framework of the MUlti-SIte COntinuous Spectroscopy (MUSICOS) campaign in 1998.}}
    \subtitle{The effect of magnetic activity in the short and long term}

   \author{L. Kriskovics
          \inst{1,2}
          \and
          Zs. K\H{o}v\'ari
          \inst{1,2}
          \and
          B. Seli
          \inst{1,2,3}
          \and
          K. Oláh
          \inst{1,2}
          \and
          K. Vida
          \inst{1,2,3}
          \and
          G. W. Henry
          \inst{4}
          \and
          T. Granzer
          \inst{5}
          \and
          A. Görgei
          \inst{1,2,3}
          } 

   \institute{Konkoly Observatory, Research Centre for Astronomy and Earth Sciences, Konkoly Thege \'ut 15-17., H-1121, Budapest, Hungary\\
              \email{kriskovics.levente@csfk.org}
        \and
            CSFK, MTA Centre of Excellence, Budapest, Konkoly Thege Mikl\'os út 15-17., H-1121, Hungary
         \and
             Eötvös Loránd University, Budapest, Hungary
         \and
             Center of Excellence in Information Systems, Tennessee State University, Nashville, TN 37209, USA
         \and
            Leibniz-Institut für Astrophysik Potsdam (AIP), An der Sternwarte 16, D14482 Potsdam, Germany
             }

   \date{Received ...; accepted ...}

  \abstract
   {Homogeneous photometric time series spanning decades provide a unique opportunity to study the long-term cyclic behaviour of active spotted stars such as our target EI\,Eridani. In addition, with ultra-precise space photometry data, it is possible to investigate the accompanying flare activity in detail. However, the rotation period of $\approx$2 days for EI\,Eri makes it impossible to achieve time-resolved surface images from a single ground-based observing site. Therefore, for this purpose, spectroscopic data from a multi-site observing campaign are needed.}
   {We use our photometric time series of more than forty years to analyze the long-term behaviour of EI\,Eri. Flare activity is investigated using space-borne photometric data obtained with the Transiting Exoplanet Survey Satellite (TESS). The MUlti-SIte COntinuous Spectroscopy (MUSICOS) campaign in 1998 aimed to achieve high-resolution, multi-wavelength spectroscopic observations from many sites around the globe, so that uninterrupted phase coverage of EI\,Eri became available. We use these data to reconstruct successive surface temperature maps of the star in order to study the changes of starspots on a very short timescale.}
   {We use long-term, seasonal period analysis of our photometric time series to study changes in the rotational period. Short-term Fourier-transform is also applied to look for activity cycle-like changes. We also study the phase and frequency distribution of hand-selected flares. We apply our multi-line Doppler imaging code to reconstruct four consecutive Doppler images. These images are also used to measure surface differential rotation by our cross-correlation technique. In addition, we carry out tests to demonstrate how Doppler imaging is affected by the fact that the data came from several different instruments with different spectral resolutions.}
   {Seasonal period analysis of the light curve reveals a smooth, significant change in period, possibly indicating the evolution of active latitudes. Temperature curves from $B-V$ and $V-I$ show slight differences, indicating the activity of EI\,Eri is spot dominated. Short-term Fourier transform reveals smoothly changing cycles between 4.5--5.5 and 8.9--11.6 years. The time-resolved spotted surface of EI\,Eri from Doppler imaging enabled us to follow the evolution of the different surface features. Cross-correlating the consecutive Doppler maps reveal surface shear of $\alpha=0.036\pm0.007$. Our tests validate our approach and show that the surface temperature distribution is adequately reconstructed by our method. The tests also indicate how accurately the cross-correlation method can reproduce the surface shear as a function of the spectral resolution.}
   {}


   \maketitle
%

\section{Introduction}

EI\,Eridani = HD 26337 (G5\,IV, $P_\mathrm{rot}\ = 1.945$ days, $V = 7\fm1$)
is a well-known, rapidly rotating ($v\sin i = 51$ km/s) active RS\,CVn type, non-eclipsing,
single-lined spectroscopic binary. It has been identified as an active star by \cite{bidelman1973} based on its Ca {\sc ii} H\&K emission, which was later confirmed by \cite{fekel1980}. Later on, \cite{fekel1982ibvs} identified it as an RS\,CVn type variable. They also detected photometric variability with an amplitude of  $V\approx 0\fm2$ and a period of roughly two days. \cite{fekel1986} derived an orbital period of 1.9472 days, while \cite{hall1987} detected a photometric period of $1.945\pm0.005$ days from $UBV$ photometry. 

Its long-term photometric variability and behavior were extensively studied in the last four decades. \cite{hall1987} reported seasonal changes in the photometric period on the order of one percent. \cite{strassmeier1989} found a seasonal change of 0.043 days in the rotational period throughout three seasons. \cite{olah2012_eieri} also reported a $\pm 2\%$ deviation from the orbital period on a three-decade-long dataset. 
\begin{figure*}[h!]
\begin{multicols}{2}
    \includegraphics[width=\linewidth]{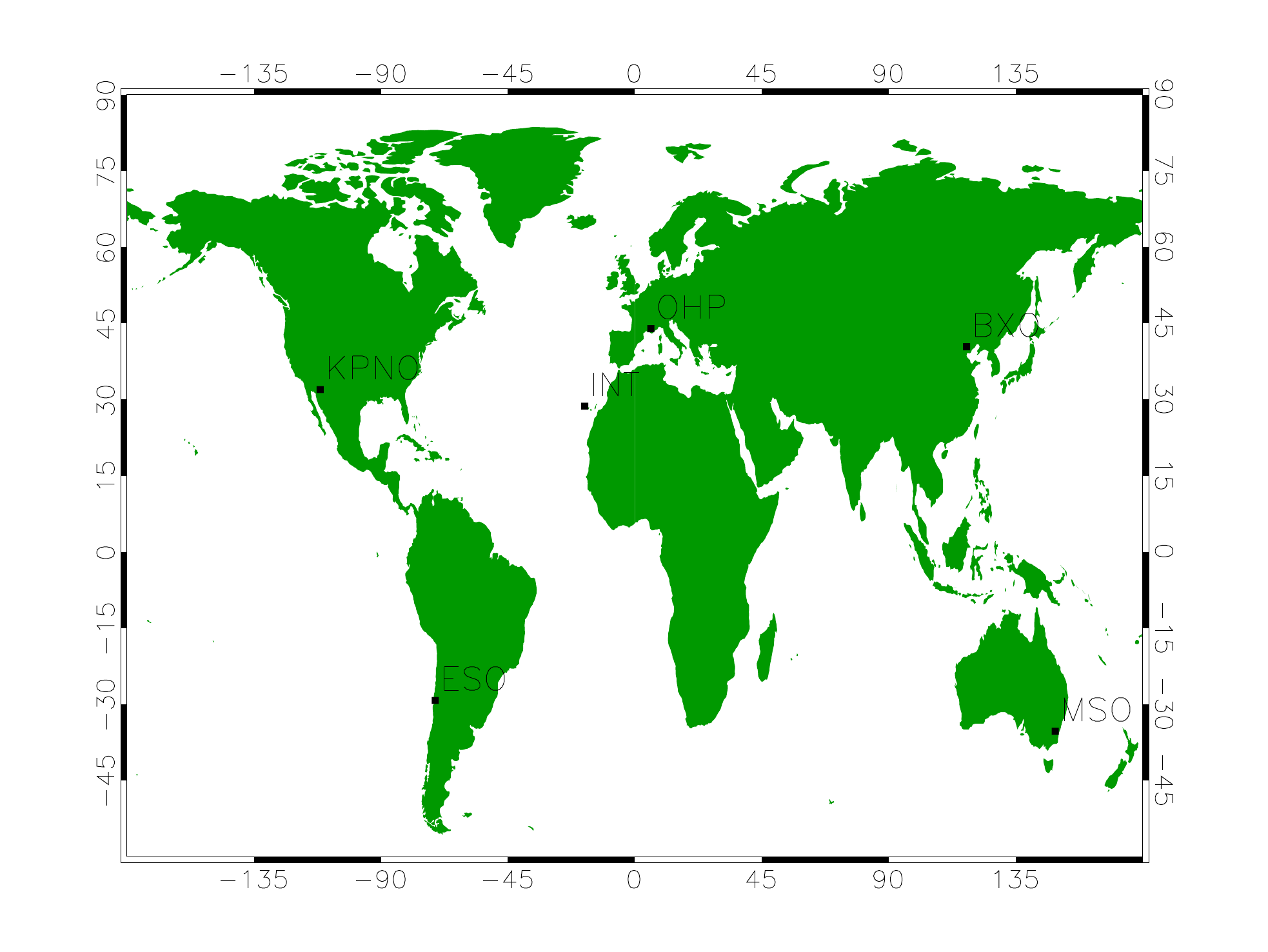}\par 
    \includegraphics[width=\linewidth]{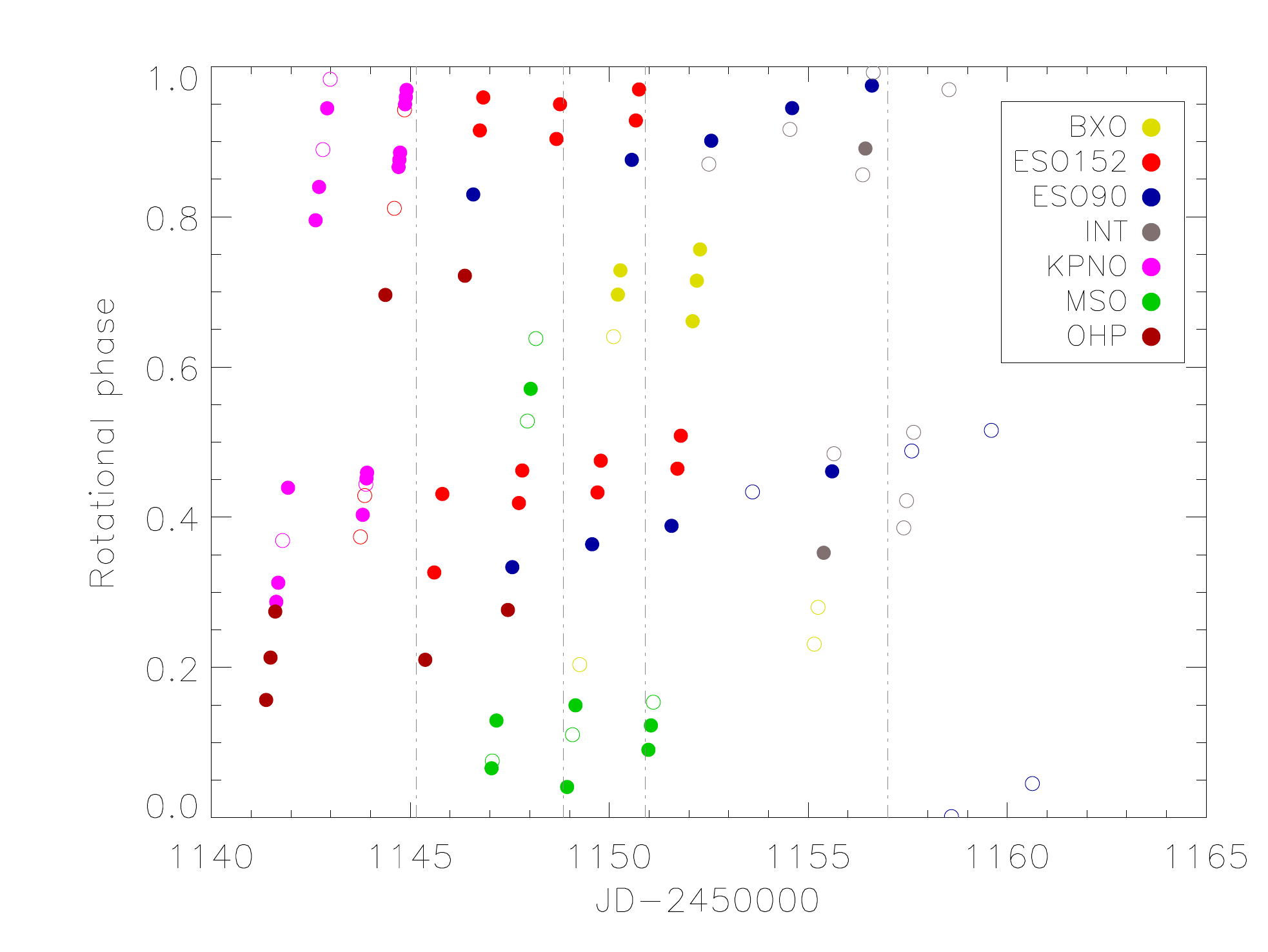}\par 
\end{multicols}
\caption{Observing sites and phase distribution of the data. \textit{Left panel: }MUSICOS sites involved in this work, namely Observatoire Haute-Provence (OHP), Xinglong Observatory (BXO), Kitt Peak National Observatory (KPNO), ESO La Silla (ESO), La Palma (INT) and Mt. Stromlo Observatory (MSO). \textit{Right panel: }Phases of the observations plotted against the reduced Julian Date. Different colors denote different instruments (see legend). Observations plotted with filled circles were used for Doppler imaging, while the empty circles denote omitted observations. The vertical dashed lines separate the four subsets used for Doppler imaging.}
\label{fig:obsplots}
\end{figure*}

It was \cite{rodono1992} who first suggested a long-term cycle of roughly 10 years. \cite{kgs:bartus97} also found an $11\pm1$ years-long cycle in the sinusoidal change of a 16 years long photometric time-series, without any apparent relation between the cycle and the change of the seasonal periods. \cite{olah2002_cyc} also found photometric cycles of roughly 2.4 and 12.2 years, which were later confirmed and refined by \cite{2009A&A...501..703O} with values of $\approx 2.9-3.1$ and $\approx 14$ years. Moreover, they reported a cycle with a length of roughly 4.1--4.9 years.
EI\,Eri is also known to exhibit flares. \cite{pandey2012} estimates the peak energy during flares in the 0.3--10 keV energy band to be $\approx10^{31}-10^{32}\,\mathrm{erg}$ using observations obtained by the XMM-Newton X-ray observatory.

EI\,Eri has been among the prime targets of Doppler imaging in the past decades since the first application of the technique. \cite{strassmeier1990eieri}, \cite{strassmeier1991eieri} and \cite{hatzes:vogt92} reported a permanent polar spot with a radically changing appendage. The presence of the polar feature was confirmed by \cite{wasi01} and later, \cite{wasi_2} and \cite{kovari2009_eieri}. They also reported constantly changing low latitude features over 6--10 rotations (approximately two-three weeks). 
However, since the rotational period of EI\,Eri is very close to two days, covering a rotation with optimal phase coverage from a singular observing site takes around a month, but reaching adequate coverage also requires at least two weeks of time series. On images reconstructed using these datasets, shorter-lived features are suppressed and smeared out, which renders the resulting Doppler images unfeasible to show rapid changes. It also makes differential rotation measurements less reliable. As conventional, ground-based optical observing runs are regularly interrupted by daybreak, the only solution for this issue is a multi-site campaign. Consequently, EI\,Eri was chosen as one of the main targets for the MUSICOS campaign in 1998 organized at ESTEC/ESA by Bernard Foing and Joana Oliveira \citep[for general details on the campaign, see][]{musicos88}.  MUSICOS stands for MUlti-SIte COntinuous Spectroscopy and aims to achieve high-resolution, multi-wavelength spectroscopic observations from many sites around the globe so that uninterrupted phase coverage of selected objects can be obtained. 

The MUSICOS dataset presented in this paper was used to create preliminary Doppler images in order to measure surface differential rotation by \cite{kovari2009_eieri}. They also confirm the presence of a stable polar spot with a slightly changing shape, and short-lived spots at lower latitudes, even though the rapid changes are somewhat masked by the artifacts that could be the result of the single-line inversion approach that was used. They also derived a surface differential rotation parameter of $\alpha=\Delta\Omega/\Omega_{\mathrm{eq}}=0.0037$. This is congruent with the estimation of \cite{olah2012_eieri} based on long-term photometry.

This paper is structured as follows. After presenting the observations in Sect.~\ref{sect:obs}, we study the long-term spot activity, including photometric cycles, and the seasonal change of rotational periods of EI\,Eri using a photometric time series of unprecedented length, covering four decades (see Sect.~\ref{sect:longphot}). Photometric data from the Transiting Exoplanet Survey Satellite (TESS) are used to investigate the flaring activity in Sect.~\ref{sect:flares}. The rapid changes of the spot configuration are investigated using spectral time series from the MUSICOS 98 campaign in Sect.~\ref{sect:doppler}.  We apply a multi-line Doppler imaging code on the MUSICOS dataset to reconstruct four Doppler images of EI\,Eri in order to investigate the short-term changes of the spotted surface. In the Appendix Sect.~\ref{sect:test}, we perform tests to verify that the spectra with different spectral resolutions used for Doppler imaging do not introduce artifacts and bias the results.


\begin{figure*}[t!!!]
\centering
\includegraphics[width=0.95\linewidth]{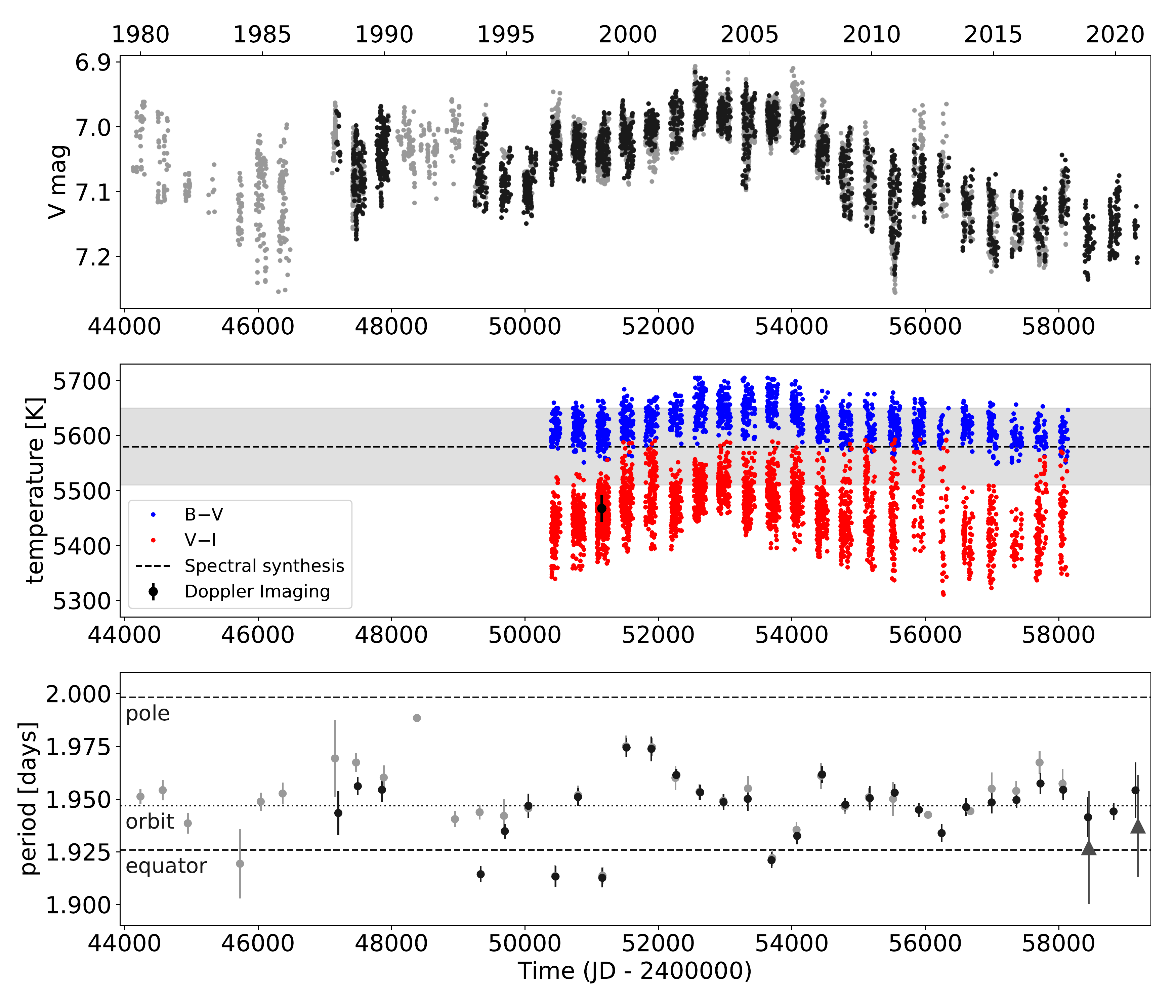}\par 
\caption{Available ground-based photometry, calculated temperature curves and yearly rotational periods for EI Eri. \textit{Top panel}: Ground-based Johnson $V$ light curve of EI\,Eri from the Potsdam Wolfgang APT at Fairborn Observatory and  from the literature (grey dots) and from \cite{2017ApJ...838..122J} and present paper (black dots). \textit{Middle panel: }Temperatures from $B-V$ (blue) and $V-I$ (red). The dashed line and the grey area denote the effective temperature from spectral synthesis with 1$\sigma$ error ($T_{\mathrm{eff}}=5580\pm70\,\mathrm{K}$, see Sect. \ref{sect:synth}). Black dot shows the averaged surface temperature value from the four Doppler maps ($\langle T\rangle=5467\pm25\,\mathrm{K}$, see Sect. \ref{sect:di}). \textit{Bottom panel:} Yearly rotational periods from the $V$ light curves with grey and black points, as on the upper panel and the two available TESS sectors (grey triangles). The dashed lines show the maximal (pole) and minimal (equator) rotational periods calculated using the differential rotation from the Doppler maps (see Sect \ref{sect:diffrot}), while the dotted line denotes the orbital period.}
\label{fig:phot}
\end{figure*}

\section{Observations}\label{sect:obs}

\subsection{Photometric observations}
Photometric observations in Johnson $V$ and $I_C$ colors were carried out with the Wolfgang--Amadeus twin Automatic Photoelectric Telescope (APT), two robotic 75-cm telescopes for photoelectric photometry at Fairborn Observatory in the Sonoran desert near Tucson, Arizona \citep{kgs:boyd97}. The observations were made differentially with HD\,25852 as the comparison star ($V = 7\fm83, V-I_C=1\fm03$) and HD\,26409 as check star ($V = 5\fm44, B-V=0\fm94$).

Johnson $B$ and $V$ data were gathered with the Tennessee State University’s T3 0.4\,m Automatic Photoelectric Telescope (APT) located also at Fairborn Observatory. For this dataset HD\,26409 was used as comparison star. Part of the $V$ dataset was published in \cite{2017ApJ...838..122J}. All photometry was transformed to the Johnson--Cousins $BV$($RI$)$_C$ system. In addition, older photometric observations since 1980 were used from the literature \citep[see][Table~5]{2009A&A...501..703O,kgs:bartus97}. The light curves of the ground-based photometry  are shown in Fig. \ref{fig:phot}.

TESS 120\,s cadence data from sectors 5 and 32 are also available for EI\,Eri and were included in the photometric analysis, see Sect.~\ref{sect:flares}.

\subsection{Spectroscopic observations}
The MUSICOS 1998 campaign involved eight northern and southern sites and ten telescopes to fulfill the needs of six scientific programs and took place from November 20 to December 14, 1998. For this work, we used data from seven instruments at six observing sites. These instruments are summarized in Table \ref{tab:sites}. Fig. \ref{fig:obsplots} shows the distribution of these observatories: Observatoire Haute-Provence (OHP), Xinglong Observatory (BXO), Kitt Peak National Observatory (KPNO), ESO La Silla (ESO), La Palma (INT) and Mt. Stromlo Observatory (MSO). 

\begin{table}[t!!!]

\begin{threeparttable}

\caption{Sites and telescopes involved in this work.} 
\begin{tabular}{l c c c}  
\noalign{\smallskip}
\hline\hline
Site & Telescope & Instrument & Resolving Power\\
\hline
OHP      & 1.93\,m     & ELODIE     & 43\,000 \\
ESO90    & 0.9\,m      & HEROS      & 20\,000 \\
ESO152   & 1.52\,m     & FEROS      & 48\,000 \\
KPNO     & 0.9\,m      & Echelle    & 65\,000 \\
BXO      & 2.2\,m      & Echelle    & 35\,000 \\
MSO      & 1.9\,m      & Echelle    & 35\,000 \\
INT      & 2.5\,m      & MUSICOS    & 35\,000 \\
\noalign{\smallskip}
\hline
\noalign{\smallskip}
\end{tabular}
\label{tab:sites}

\begin{tablenotes}
\item \small Telescope abbreviations are as follows: the 1.93-m telescope at Observatoire Haute-Provence (OHP), the 90-cm and 152-cm telescopes at ESO La Silla (ESO90 and ESO152 respectively), the McMath-Pierce telescope at Kitt Peak National Observatory, Xinglong Observatory, Beijing (BXO),  Mt. Stromlo, Australia (MSO) and the Isaac Newton Telescope at La Palma (INT).
\end{tablenotes}
\end{threeparttable}
\end{table}

For this specific program, a total of 90 high-resolution spectra of
EI\,Eri were obtained within a time of 21 days (November 23 -- December 13). The observing log is summarized in Table \ref{tab:obslog}. The various observing sites made use of different spectrographs (long slit, echelle), which leads to data sets of different quality, different wavelength ranges, and spectral resolutions. Fortunately, the region around 6400--6450\AA, which contains frequently used Doppler imaging lines (Fe\,{\sc i}\,6400, Fe\,{\sc i}\,6408, Fe\,{\sc i}\,6421, Fe\,{\sc i}\, 6430 and Ca\,{\sc i}\,6439), is present in all of them. Several spectra were discarded due to low signal-to-noise or contaminated line profiles, resulting 59 selected spectra. 
i
The data were reduced using standard reduction techniques for echelle spectra in the NOAO packages of IRAF: bias subtraction, flat-field correction to remove pixel-to-pixel variations, and the curvature of the blaze function. For OHP, HEROS and Xinglong observations, the MIDAS package was used. Background subtraction and flat-field correction using exposures of a tungsten lamp were applied. Wavelength calibration was performed by taking the spectra of a Thorium--Argon lamp. The spectra were normalized by a low-order polynomial fit or a cubic spline fit to the observed continuum. 

Since the dataset is comprised of spectra from seven different instruments with different spectral resolutions, which can cause problems during the Doppler inversions, we decided to decrease the resolution of all spectra to the lowest available resolution of 20\,000. In Appendix~\ref{sect:test} we present a series of tests to validate this approach.

The observations are phased with the following ephemeris \citep{kovari2009_eieri}:

\begin{equation}
    \label{eq1}
    HJD = 2448054.7130 + 1.9472287 \times E.
\end{equation}

Table \ref{tab:obslog} shows the observing log, while Fig.~\ref{fig:obsplots} shows the phase distribution of the different measurements.

\begin{table}[b!!!]
\begin{center}
\caption{Astrophysical parameters from spectral synthesis.}
\begin{tabular}{lc}
\noalign{\smallskip}
\hline
\noalign{\smallskip}
$T_{\mathrm{eff}}$ & $5580 \pm 70$ K\\
$\log g$ & $3.75 \pm 0.35$ \\
$[\mathrm{Fe/H}]$ & $-0.27 \pm 0.18$ \\
$v_{\mathrm{mic}}$ & $2.0 \pm 0.5$ km/s\\
$v\sin i$ & $51\pm1$ km/s \\
\hline
\noalign{\smallskip}
\end{tabular}

\label{tab:aphyspar}
\end{center}
\end{table}

\section{Photometric analysis}\label{subsect_phot}

\subsection{Analysis of the long-term photometry}\label{sect:longphot}

\begin{figure}[t!!!]
    \includegraphics[width=\linewidth]{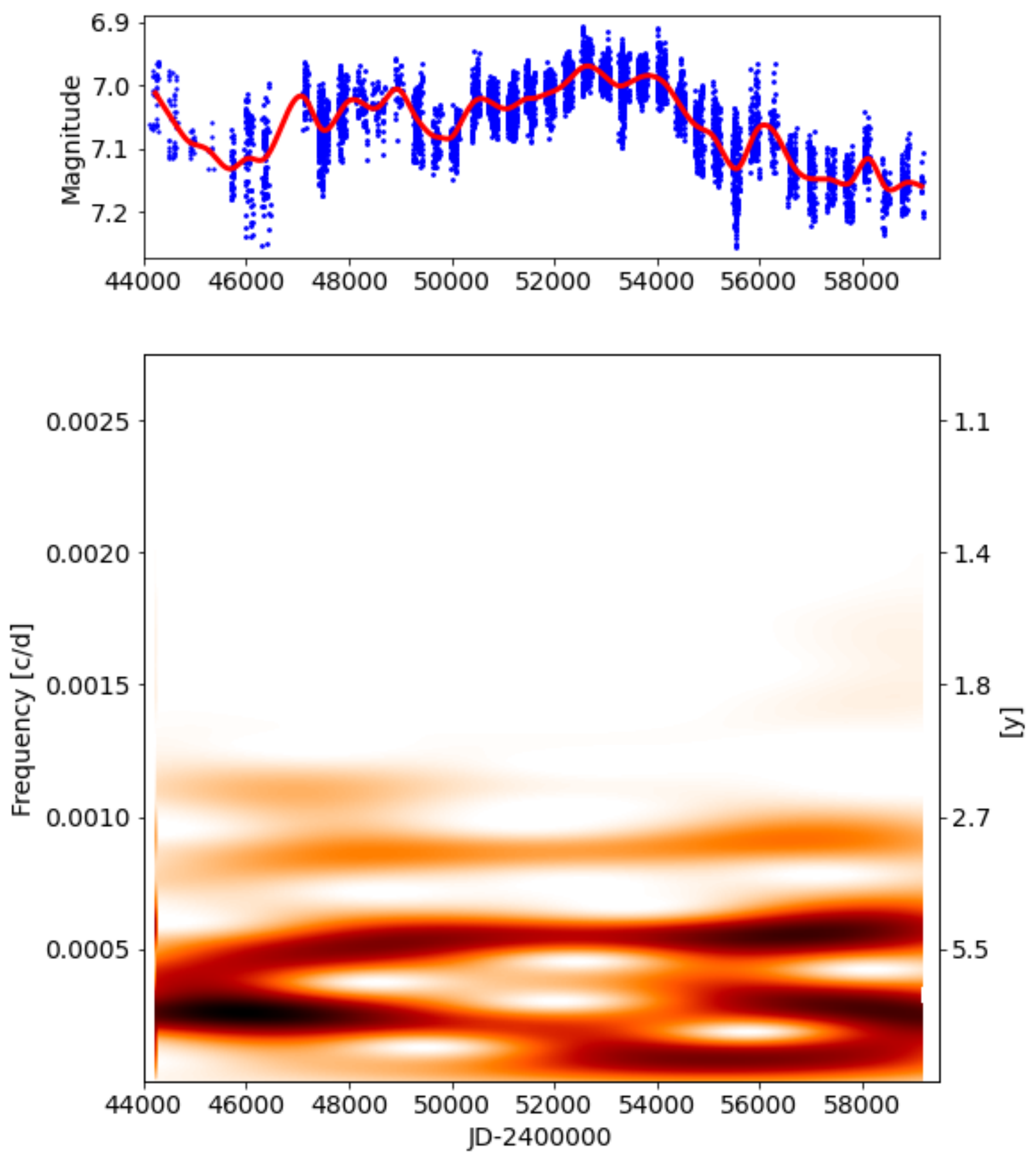}\par 
\caption{Short-term Fourier transform of all available $V$ observations of EI\,Eri. A smoothly changing cycle is seen between 4.5-5.5 years (and its half), and a longer one which moves between 8.9-11.6 years.}
\label{fig:stft}
\end{figure}

\begin{figure*}[t!!!]
\centering
    \includegraphics[width=0.44\linewidth]{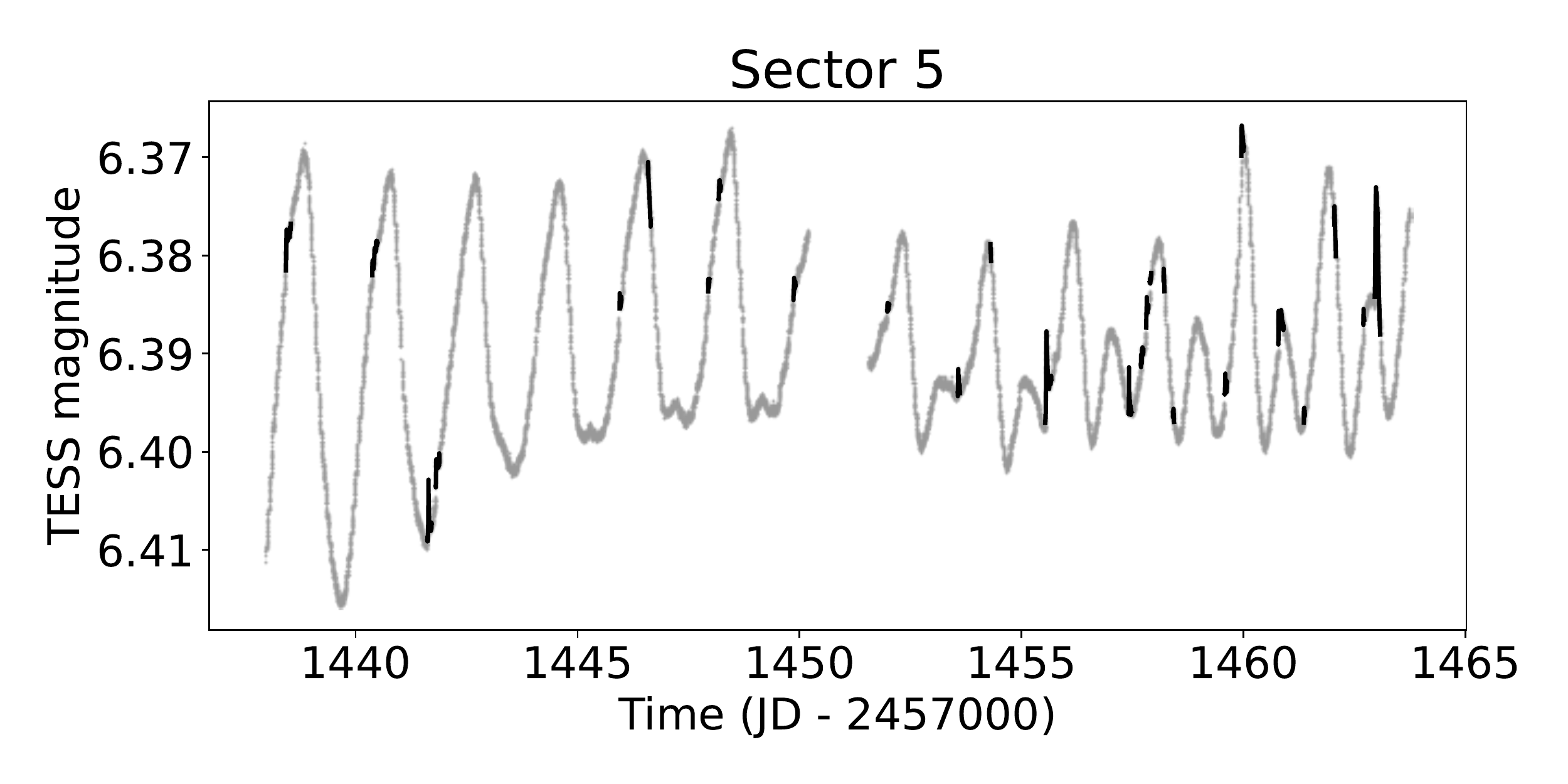}
    \includegraphics[width=0.44\linewidth]{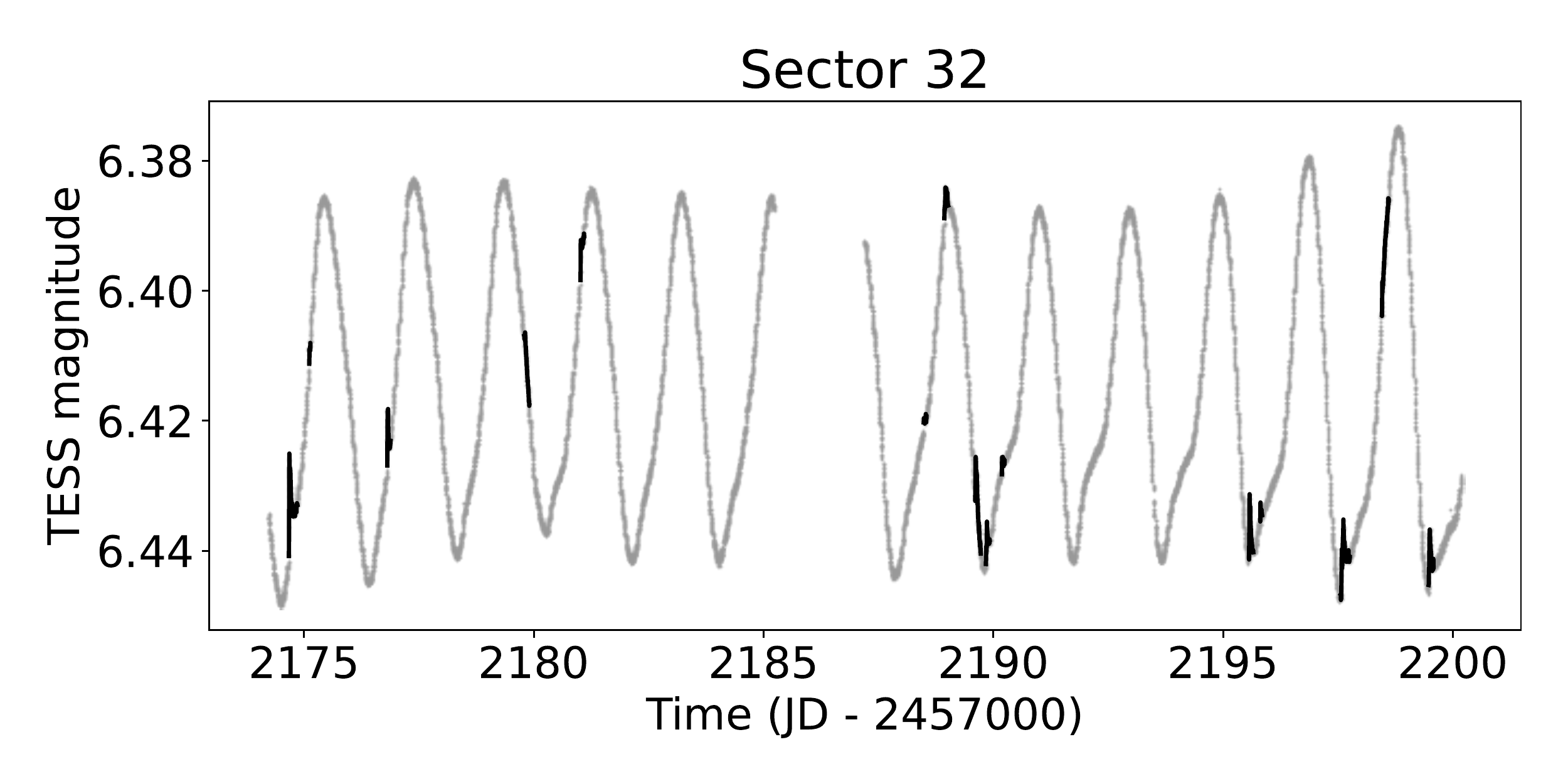}
\caption{TESS observations from Sector~5 (left) and Sector~32 (right). The identified flares are marked with black.}
\label{fig:flare}
\end{figure*}

Due to differential rotation, the period measured from photometry can slightly change, depending on the latitude where the dominant spots reside. Fig-.~\ref{fig:phot} shows the measured rotational period of EI\,Eri in seasonal blocks, using the peak of the Lomb-Scargle periodogram \citep{Lomb,Scargle} with uncertainties calculated from the width of the peaks at 90\% power. The periods from the literature and Potsdam APT data (grey dots) are consistent with the data from \cite{2017ApJ...838..122J} and present paper (black dots) and with the two points from the TESS light curves (grey triangles) within one season. There is a smooth and significant change in the period, indicating the evolution of active latitudes.

The middle panel of Fig.~\ref{fig:phot} shows the temperatures calculated from $B-V$ and $V-I$ color indices by linearly interpolating the grid of \cite{Worthey2011} at the given $\log g$ and metallicity. The dashed line and the grey area denote $T_{\mathrm{eff}}=5580\pm70\,\mathrm{K}$ from spectral synthesis with $1\sigma$ error (for details, see Sect. \ref{sect:synth}). The average temperature calculated from the Doppler maps is plotted with a black dot, and the $1\sigma$ error is computed from the four different values for the four different maps (for details on the inversion, see Sect. \ref{sect:di}).

To look for activity cycles, we used Short-term Fourier transform \citep
{2009A&A...501..695K} for the present dataset, which is 41 years long, 13 years longer than the one used in \cite{2009A&A...501..703O}. Since the algorithm requires uniform sampling in time, a cubic spline interpolation was used. We find that the previously identified structures on the STFT diagram are still present, indicating roughly the same cyclic behaviour. The signal of the longest, high-amplitude feature in Fig.~\ref{fig:stft} is suppressed, but above about 13.6 years, which is one-third of the length of the data, all signals have the same amplification. Smoothly changing cycles are seen between 4.5--5.5, and 8.9--11.6 years, which are not harmonics, since the speed and the direction of their changes are different. A weak cycle-like feature of about 2.5 years is also seen at the beginning of the time span.

\subsection{Flares}\label{sect:flares}
\begin{figure}
\includegraphics[width=\linewidth]{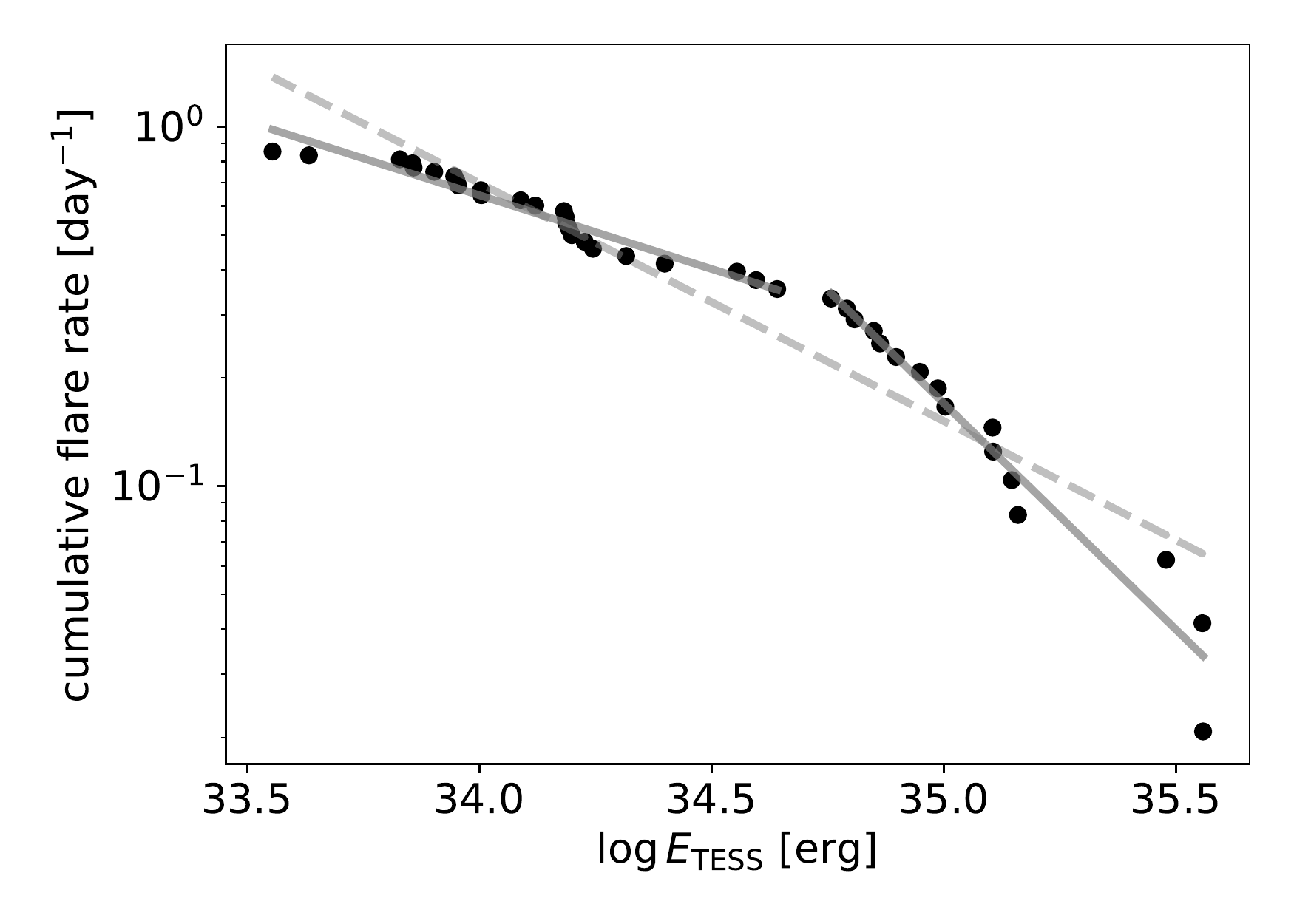}
\caption{Flare frequency distribution from the TESS data. The dashed line denotes a fit for all of the points with a power-law index of $1.66\pm0.04$, while the two solid lines show a two-component fit ($1.41\pm0.02$ and $2.26\pm0.08$), see Sect.~\ref{sect:flares}.}
\label{fig:flare_FFD}
\end{figure}

We manually identified 41 flares in the two available TESS sectors (see Fig.~\ref{fig:flare}). To calculate their energies in the TESS band, we followed similar procedures as in \cite{flaring_giants}, using a BT-NextGen model spectrum \citep{1999ApJ...512..377H} convolved with the TESS transmission curve to get the quiescent stellar luminosity of $L_{\rm TESS} = 4.69 \cdot 10^{33} {\rm erg\,s}^{-1}$, and integrating over the flare light curves to get the equivalent durations. The flare frequency distribution (FFD) in Fig.~\ref{fig:flare_FFD} shows a broken power law shape. Two-component fits of the FFD yield  power law indices of $1.41\pm0.02$ and $2.26\pm0.08$ (one minus the slope of the fitted line) for the first and second parts, respectively, while a fit for all datapoints gives $1.66\pm0.04$ (for further discussion, see Sect.~\ref{sect:disc}). 

\begin{figure}
\includegraphics[width=\linewidth]{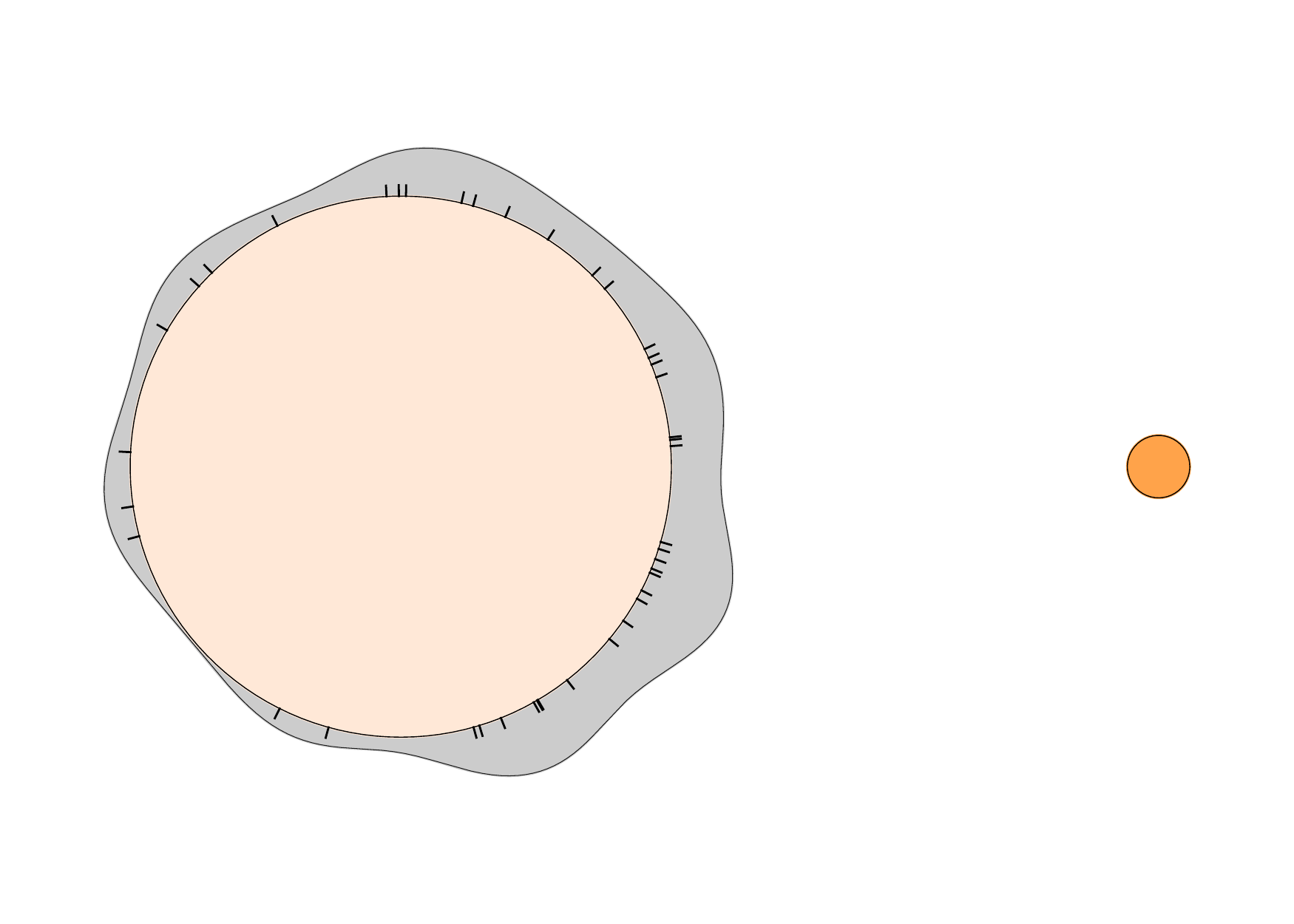}
\caption{Phase distribution of the TESS flares on the primary component in the reference frame of the binary orbit. The relative radii and separation of the circles are for scale. The colors are realistic digital colors for the given spectral types from \cite{stellar_RGB_colors}. The grey area shows a Gaussian kernel density estimation with the bandwidth of $2^\circ$. Black ticks denote the orbital phase values of the flare observations.}
\label{fig:flare_phases}
\end{figure}

Fig.~\ref{fig:flare_phases} shows the phase distribution of flares, in the reference frame of the orbit, using the ephemeris from Eq.~\ref{eq1}. While the spot configuration of EI Eri changes quickly, the orbiting secondary component provides a solid reference frame. To test whether there is a significant increase of flares at given phases, we run a two-sided Kuiper's test \citep{kuiper1960tests}. It compares the measured data to a given distribution, similar to the Kolmogorov--Smirnov test, but it is invariant under a cyclic transformation, which makes it applicable to compare cyclic phase distributions (the Kolmogorov--Smirnov test would give different results for different starting epochs). The test rejects the null hypothesis of a uniform flare phase distribution with a $p$-value of 0.006. The increase in flare rate appears on the side facing the secondary component.

\section{Doppler imaging}\label{sect:doppler}

\subsection{Astrophysical parameters for the inversion}
\label{sect:synth}
Precise astrophysical parameters are fundamental for Doppler inversion. Therefore we carried out a detailed spectroscopic analysis based on spectral synthesis using the code SME \citep{piskunov_sme} and MARCS atmospheric models \citep{gustafsson_marcs}. Atomic line parameters were taken from the VALD database \citep{kupka_vald}. Macroturbulence was estimated using the following equation \citep{valenti_macro}:

\begin{equation}
    v_{\mathrm{mac}}=\Bigg(3.98-\frac{T_{\mathrm{eff}}-5770 \mathrm{K}}{650 \mathrm{K}}\Bigg)\,\mathrm{km\,s}^{-1}.
\end{equation}

Astrophysical parameters were fitted one-by-one iteratively as described in \cite{kriskovics2019}. The results are summarized in Table \ref{tab:aphyspar}.
\begin{figure*}[t!!!]
\vspace{2.5cm}\Large{DI1}

\vspace{-2.5cm}\hspace{1.3cm}\includegraphics[width=0.93\linewidth]{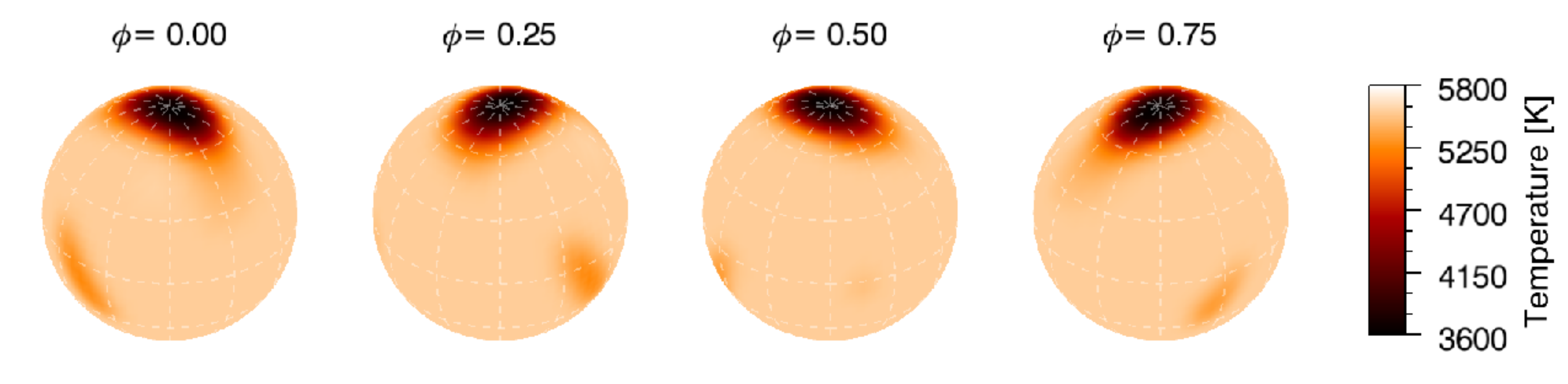}\par
\vspace{2.5cm}\Large{DI2}

\vspace{-2.5cm}\hspace{1.3cm}\includegraphics[width=0.93\linewidth]{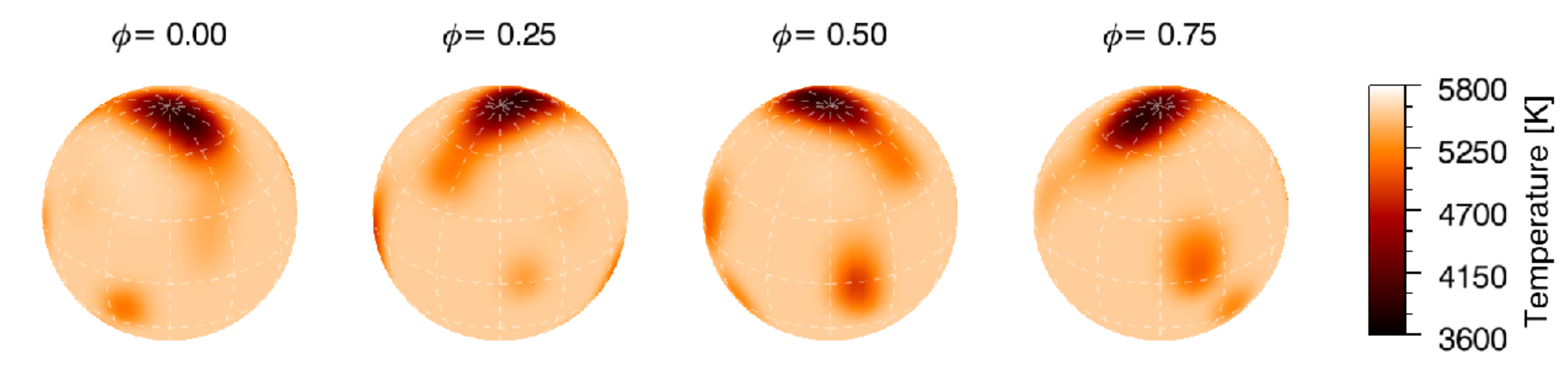}\par
\vspace{2.5cm}\Large{DI3}

\vspace{-2.5cm}\hspace{1.3cm}\includegraphics[width=0.93\linewidth]{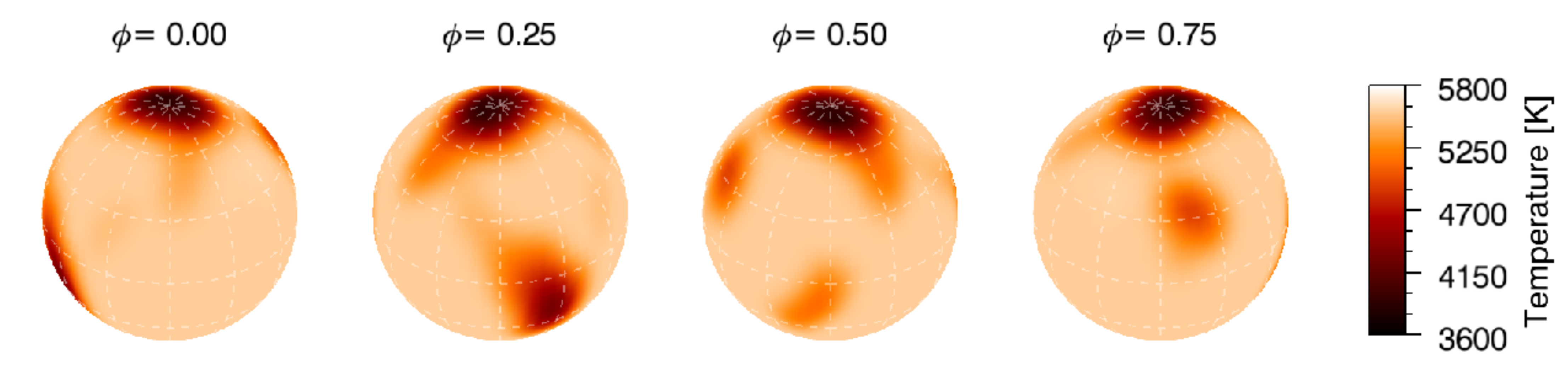}\par
\vspace{2.5cm}\Large{DI4}

\vspace{-2.5cm}\hspace{1.3cm}\includegraphics[width=0.93\linewidth]{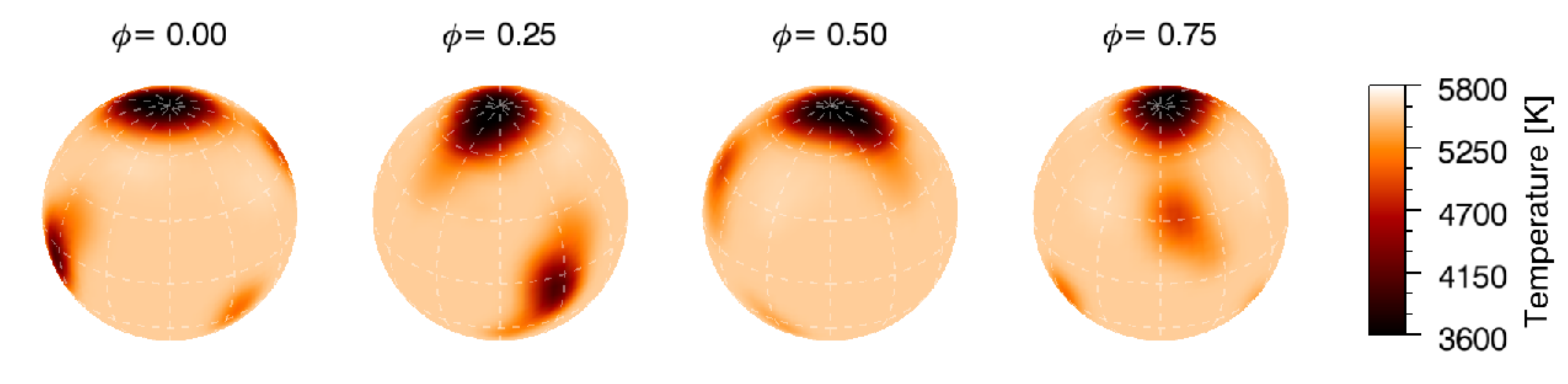}\par
\caption{Four consecutive Doppler images of EI\,Eri obtained for the MUSICOS 1998 data. The corresponding time intervals for the four images from top to bottom are 2451141.3758--2451144.9045, 2451145.6007--2451148.7618, 2451148.9391--2451150.7474 and 2451150.9827--2451156.5520.}
\label{fig:di_eieri}
\end{figure*}
\subsection{The Doppler imaging code iMap}
The Doppler imaging code \textit{iMap} \citep{carroll_imap}, which is used in this work, carries out multi-line Doppler inversion on a list of photospheric lines.
Since the wavelength coverage of our dataset is limited, we used five (Fe\,{\sc i}\,6400, Fe\,{\sc i}\,6408, Fe\,{\sc i}\,6421, Fe\,{\sc i}\, 6430 and Ca\,{\sc i}\,6439) virtually unblended lines with well-defined continuum, suitable line depth, and temperature sensitivity. The stellar surface was divided into $6^{\circ}\times 6^{\circ}$ segments. For each local line profile, the code utilizes a full radiative solver \citep{carroll_solver}.

Then the local line profiles are disk integrated, and the individually modeled,  disk-integrated lines are averaged. Atomic line data are taken from VALD \citep{kupka_vald}. Model atmospheres are taken from \cite{castelli_mod}
and are interpolated for the necessary temperature, gravity, or metallicity values. Due to the high computational capacity requirements, LTE radiative transfer is used instead of spherical model atmospheres, but imperfections in the fitted line shapes are well compensated by the multi-line approach.  Additional input parameters are micro- and macroturbulence, and $v\!\sin\!i$. Table \ref{tab:aphyspar} summarizes the astrophysical parameters used during the inversion. Inclination was set to $i=56^{\circ}$ \citep[adopted from][]{wasi_2}.

For the surface reconstruction, an iterative regularization method based on a Landweber algorithm is used \citep{carroll_imap}, meaning no additional constraints are imposed in the image domain. 

\subsection{Time-resolved surface evolution}
\label{sect:di}
The 59 selected spectra were divided into four subsets (hereafter DI1, DI2, DI3 and DI4) corresponding to the following Julian Date intervals: 2451141.3758--2451144.9045, 2451145.6007--2451148.7618, 2451148.9391--2451150.7474 and 2451150.9827--2451156.5520, with lengths of roughly 1.8, 1.6, 1.0 and 2.9 rotations, respectively. Imaging lines were selected from the spectral region that was covered by all of the spectra in order to make the resulting maps comparable. This gives us a unique opportunity to study the short-term spot evolution of EI\,Eri. The selected spectral lines are Fe\,{\sc i}\,6400, Fe\,{\sc i}\,6408, Fe\,{\sc i}\,6421, Fe\,{\sc i}\, 6430 and Ca\,{\sc i}\,6439.

The resulting four Doppler images (DI1--DI4) are shown in Fig.~\ref{fig:di_eieri}. The corresponding line profile fits are shown in Fig.~\ref{fig:di_eieri_profs}, with corresponding root mean square deviations of 0.0030, 0.0030, 0.0036 and 0.0040, respectively.  The four maps show a persistent polar feature (with a spot temperature of $\Delta T \approx 1100\,\mathrm{K}$ below the temperature of the unspotted surface.
(Here we refer to our tests presented in Sect.~\ref{sect:test} of the Appendix, during which we also cover the accuracy of the spot temperature determination.) 
Associated with the dominant polar spot, a less prominent appendage appears in DI2 around $\phi$ $\approx$0.35 phase value, which is further strengthened in DI3. In DI4, the appendage starts to fade, but it is still present. The overall contrast of the polar feature seems to increase from DI1 towards DI4. With all this, several low-latitude features show rapid evolution as well. 
In DI1, the barely visible spot at $\phi\approx$0.45 in the lower hemisphere becomes stronger in DI2, and after a slight retrograde shift seen in DI3, it disappears completely in DI4, being present for $\sim4.5$ rotations. Around 0.2 phase, a low latitude feature emerges in DI2 and becomes more and more prominent.
In DI2 at $\phi\approx0.7$, a low latitude spot forms and strengthens along DI3 and DI4, while gradually shifting toward the visible pole. 

\begin{figure*}[t!!!]
\begin{multicols}{2}
    \includegraphics[width=\linewidth]{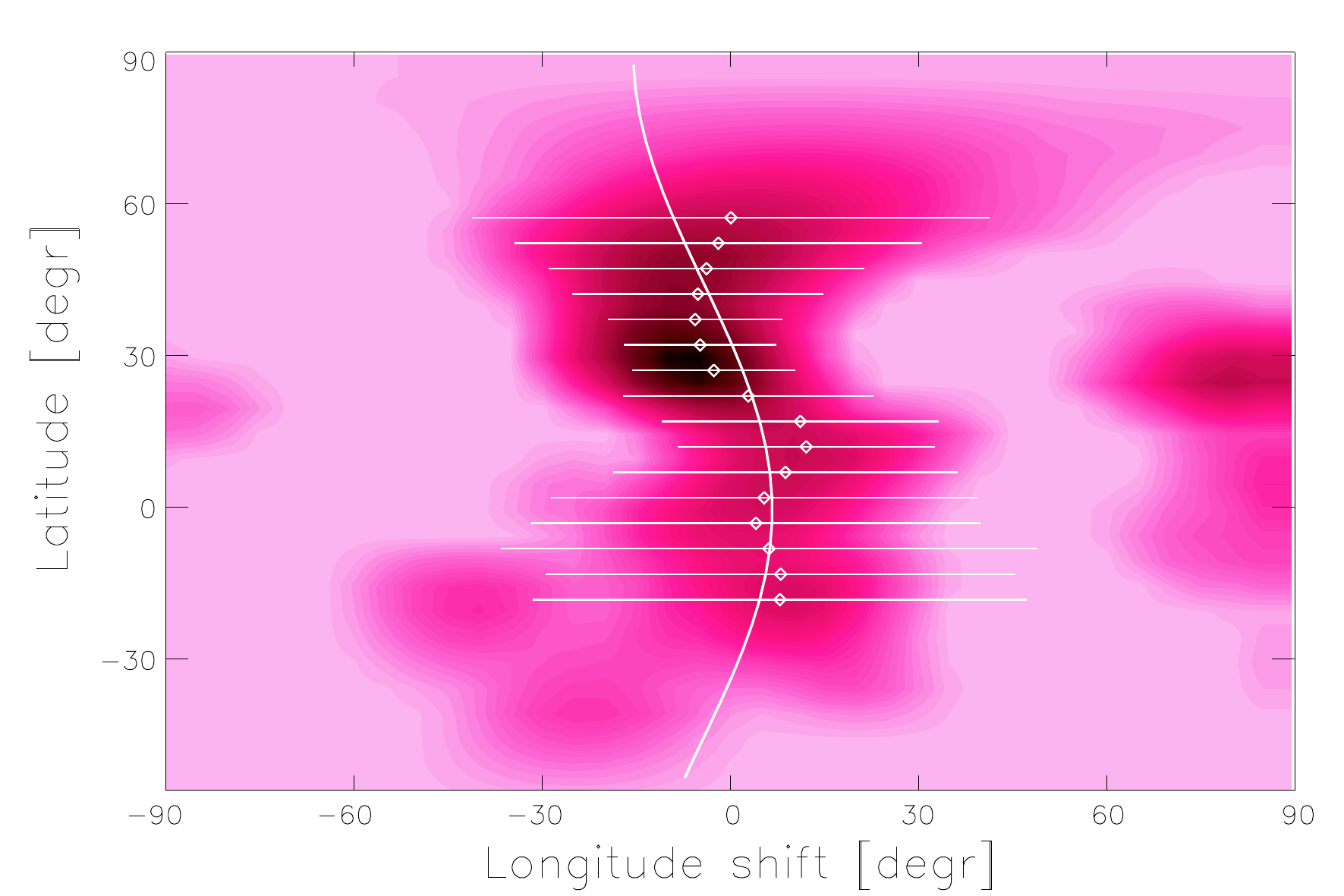}\par 
    \includegraphics[width=\linewidth]{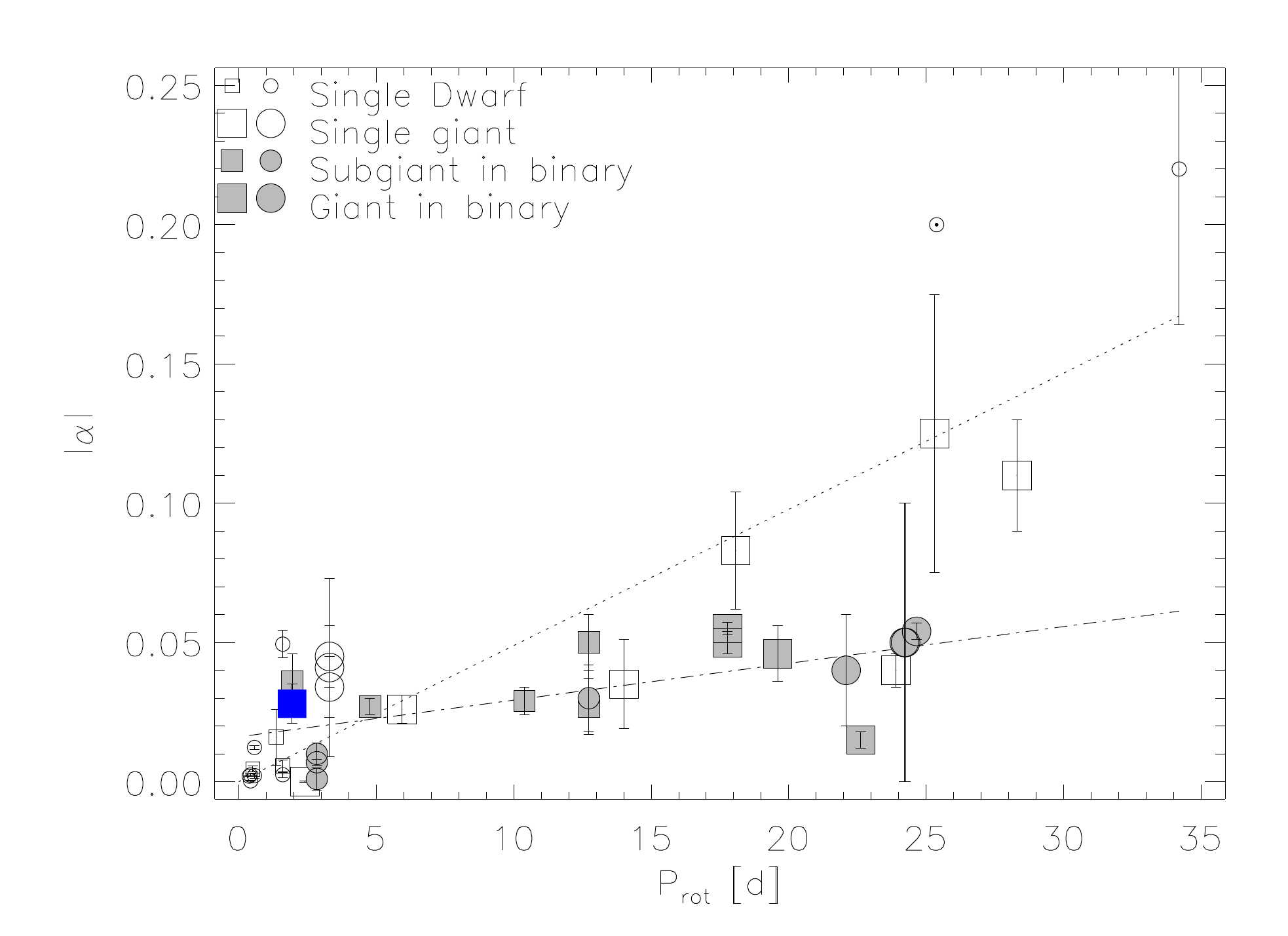}\par 
\end{multicols}
\caption{Measured surface differential rotation of EI Eri and its comparison to other stars. \textit{Left panel: } Average cross-correlation map for EI\,Eri. The correlation peaks (white circles) are fitted with a quadratic differential rotation law. The resulting fit (continuous line) indicates a solar type differential rotation with a shear parameter $\alpha=0.036\pm0.007$. \textit{Right panel:} An extended version of the surface shear coefficient versus rotational period plot from \citealt{kovari_diffrot}. Squares denote results from the cross-correlation technique, while circles show differential rotational coefficients from the sheared image method. White symbols correspond to single, greys to binary stars. Symbol size increases from dwarfs to subgiants to giants. EI\,Eri is represented by the blue square at $P_{\rm rot}\approx$2 days. The Sun is plotted with a dotted circle. The dotted and dash-dotted lines denote the linear fits to the points corresponding to singles and binaries, respectively, with slopes of $\lvert\alpha\rvert \propto (0.0049\pm0.00001) P_{\mathrm{rot}}$ and $\lvert\alpha\rvert \propto (0.0014 \pm 0.00003) P_{\mathrm{rot}}$. }
\label{fig:ccf_eieri}
\end{figure*}

In the end, it is worth emphasizing that the evolution of the different surface features can be followed quite nicely through the four \textit{independently} reconstructed maps. Therefore we believe that most of the recovered surface features are real and not imaging artifacts.

\subsection{Differential rotation}
\label{sect:diffrot}

A usual way of measuring surface differential rotation is cross-correlating consecutive Doppler images \citep{donati_ccf}. However, rapid spot evolution can easily hinder the correlation pattern, e.g., through the formation or dimming of spots or interaction between nearby spots, all of which can be seen in the Doppler images of EI\,Eri. To overcome this, and strengthen the signal of differential rotation, we use our code ACCORD \citep[e.g][]{kovari2012}, which uses the average of the cross-correlations of consecutive Doppler images.
The latitudinal correlation peaks in the resulting average correlation map are fitted with a quadratic rotational law in the form 
\begin{equation}
    \Omega(\beta)=\Omega_{\mathrm{eq}}-\Delta\Omega\sin^2\beta,
\end{equation}
where $\Omega(\beta)$ is the angular velocity at $\beta$ latitude, $\Omega_{\mathrm{eq}}$ is the angular velocity of the equator, and $\Delta\Omega=\Omega_{\mathrm{eq}}-\Omega_{\mathrm{pole}}$ gives the difference between the equatorial and polar angular velocities. With these, the dimensionless surface shear parameter $\alpha$ is defined as $\alpha=\Delta\Omega/\Omega_{\mathrm{eq}}$.

For the average cross-correlation we formed three pairs of consecutive Doppler images, that is DI1-DI2, DI2-DI3, and DI3-DI4. After the cross-correlation of the image pairs, we normalized each resulting correlation map to the same time difference so that they can be averaged. For a more detailed description of the ACCORD method see \cite{kovari_sgem} and their references. In the left panel of Fig.~\ref{fig:ccf_eieri} the average correlation pattern is fitted with a quadratic differential rotation law which yielded $\alpha=0.036\pm0.007$ ($\Omega_{\mathrm{eq}}=186.922\pm0.384$ degree/day, $\Delta\Omega=6.768\pm1.349$ degree/day). The errors are estimated from the amplitudes and the full width at half maximum values (FWHMs) of the Gaussian functions fitted to each longitudinal strip.

In the right panel of Fig.~\ref{fig:ccf_eieri} we put the surface shear of EI\,Eri against the rotation period together with other well-known active stars as an updated plot of \cite{kovari_diffrot}. Our resulting shear value of 3.6\% for EI\,Eri is in good agreement with the value expected according to the plot.
We note, however, that the differential rotation pattern can be somewhat diminished by the relatively low spectral resolution; see our tests to investigate this question in Appendix~\ref{sect:diffrottest}. 

\section{Discussion}
\label{sect:disc}

\subsection{Long term photometric behaviour}

\cite{2009A&A...501..695K} found shorter cycles of 2.9--3.1 and 4.1--4.9 years and a longer cycle of roughly 14 years on EI\,Eri. Our cycle lengths derived from a 13 years longer dataset of roughly 2.5, 4.5--5.5 and 8.9--11.6 years are in good agreement with these previous findings (in Fig.~\ref{fig:stft}, the longest 13.6 year cycle is suppressed since the dataset is only three times longer than this value). The cycle lengths are smoothly changing, which is a well-documented behavior of activity cycles, see examples in e.g. \cite{2009A&A...501..703O}, or the solar case in \cite{2009A&A...501..695K}. Visual inspection of the curve of seasonal rotation period changes (bottom panel of Fig. \ref{fig:phot}) hints at a cycle-like behavior of around 3500--4000 days (roughly 9.5--11 years). 
The similarity between the lengths of the cycle-like features of the long-term photometry and the change of the seasonal rotational periods around 9--11 years is indeed remarkable. However, there are problems in interpreting the photometric results due to the huge spots and the inclination of the stellar disk to the line of sight, which makes the tracing of the spot's position and sizes complicated. Therefore, one-by-one correspondence of the brightness changes and rotational periods (like a butterfly diagram) cannot be drawn.

The temperatures calculated from $B-V$ and $V-I$ color indices show slightly different amplitudes. The higher temperature of the facular/plage regions contributes to the signal in the $B$ bandpass, although the spots from the photosphere also have a lower contribution in $B$,
whereas the signal in the $I$ filter has no or negligible contribution from the higher temperature areas. The spot--plage ratio can be different star by star, and also within spot cycles of the same star.  
The difference between the $B-V$ and $V-I$ color indices could indicate that the activity of EI\,Eri is dominated by spots, rather than plages. 
This is also supported by the large spotted areas present on the Doppler maps and the fact that the temperature from $B-V$ correlates with the effective temperature from the spectral synthesis, while the average temperature from the Doppler images is in the interval of the temperatures calculated from $V-I$. 
We note, however, that the photometric accuracy of the color indices may hinder this result. 
Uncertainties in the order of 0.01 magnitude translate to an error of about 30~K in temperature. Nevertheless, average temperature coming from summing up the temperature elements of the surface of the Doppler images agree very well with the temperature values of the same season resulting from $V-I$, which indicates a reasonable accuracy of the results from the color indices.

\subsection{Flare activity}

The possible origin of breakpoints in FFDs was discussed by \cite{2018ApJ...854...14M} for flare stars showing a critical energy value on the FFD when the flaring loop size exceeds the local scale height, which depend on local field strengths and densities. Above this value, the flares have higher energy and the power law index is steeper, and the opposite stands below this energy value for smaller loops.

Extrapolating their results, \cite{2021A&A...647A..62O} gave examples of broken power-law FFDs for two giant stars with different breakpoints of their FFDs  but similar energy ranges suggesting different atmospheric properties and strengths of magnetic fields. In the case of the subgiant EI\,Eri this rationale may hold as well. The median flare duration of about 1.5 hours and the energy range shown in Fig.~\ref{fig:flare_FFD} are fully consistent with the general picture of flare energy--duration diagram given by \citet[][see their Fig.~14]{2021PASJ...73...44M}, placing EI\,Eri somewhere between G dwarfs and giants on the flare energy-flare duration diagram. 

When an active star is in a close binary system, the tidal effects may play a role in the position of the emerging magnetic flux tubes that are supposed to be the origin of activity on stars. Such effects supposing different physical circumstances (like the strength of the magnetic field, the depth of the convection zone where the flux tubes originate, local density, etc.) were investigated in two papers by \cite{2003A&A...405..291H, holzandschuss2003} showing that clustering of flux tube emergence positions in the orbital reference frame of close binaries may appear. Our results plotted in Fig.~\ref{fig:flare_phases}, where the observed flare phases are shown in the reference frame of the orbit, show a clustering on the hemisphere facing the secondary and may originate from such an effect.

\subsection{Rapid spot evolution and differential rotation}

Rapid spot evolution is not unheard of on subgiants. \citet{strassmeier_v711tau} found fast-evolving, low latitude features on V711\,Tau, a similar binary subgiant (although we note that V711\,Tau is somewhat cooler and its secondary is more massive). They found that in several cases, low-latitude features changed shape and became cooler on a scale of a few days -- similar to what we found on EI\,Eri. Gradual strengthening of the polar feature is also observed on V711\,Tau and EI\,Eri. It might indicate spot migration towards the poles. Moreover, \citet{haru_huvir} reported rapid spot evolution on another subgiant, HU\,Vir, with a prominent, high-latitude, but not polar feature, which drastically changes shape from one rotation to the next. We note, however, that due to the nature of Doppler imaging, we have no information about the spot evolution on time scales shorter than the $\approx$10-d rotational period of HU\,Vir).

Polar features persisting for years have been previously observed on subgiants \citep[see e.g. ][and references therein]{hussain_starspot}, also it is a fairly common and well-explained phenomenon (even from a theoretical viewpoint) on fast-rotating active stars \citep[e.g][]{isik_starspot}. However, rapid spot changes on the scale of a few days are much more difficult to explain with models. \citet{isik_starspot} even point out that the lifetime of their simulated bipolar magnetic regions (BMRs) are longer on a star similar to V711\,Tau than on a solar analog. According to their simulation, a large BMR above $\beta=70^{\circ}$ latitude can be present for more than two years. This may even be consistent with the permanent polar features of EI\,Eri.

However, BMRs and spots can behave differently: BMRs consist of spots, plages, and short-lived structures. It is also debated whether large spots on Doppler images are single magnetic structures or clusters of smaller spots. Accordingly, it is possible that different evolution timescales are applicable to a homogeneous, large spot than to several smaller spots.
\citet{isik_starspot} showed that a cluster of 
spots (that still cannot be resolved by Doppler imaging) can dissipate the fastest in their simulation, especially when large-scale flows (e.g., meridional circulation) are introduced, with a decay time of at least a few tens of days.

\citet{strassmeier_v711tau} theorized that magnetic reconnection could play an important role in this phenomenon: it may induce magnetic field diffusion from the stronger flux tubes (i.e., the cooler spots) towards the weaker (warmer) ones, resulting in increased magnetic pressure and thus more suppressed convection and rapid cooling in the warmer spots. They also estimated that on V711\,Tau, the Alfvén-velocity for the photosphere region corresponding to $\tau$=1 optical depth results in a magnetic interaction timescale between spots, which is on the order of a day. Thus, energy transport by Alfvén-waves may contribute to the rapid spot changes. 

\citet{rempel2014} carried out three-dimensional numerical simulations to study spot emergence and decay. They found that in the first phase of spot decay, the process is dominated by downward convective motion, while in the second phase, plasma intrusions fragment larger spots. The latter could be consistent with the assumptions of \citet{strassmeier_v711tau}. \citet{rempel2014} pointed out that strong, subsurface large-scale convective motions can play a significant role in spot decay (however, their simulation box was limited to the upper part of the convective zone). 

\citet[][and their Fig.~6]{namekata_spot_lifetime} compares observed stellar spot lifetimes on different solar-type stars. Their result suggests that if spot evolution is driven by the same mechanism in all evolutionary stages, the typical spot lifetime on EI\,Eri should be in the range of 10--30 days. Our findings (and previous ones in the literature, see above) are in agreement with these results, since most of the active nests are present on all of the Doppler images covering altogether 15 days, although with changing shape and contrast \citep[in agreement with][]{strassmeier_v711tau}. The TESS light curves (Fig.~\ref{fig:phot}) also support this, e.g. a new nest seems to form around JD=2457143 which is persistent for at least $\approx$10 days, although with different amplitude.
We note, however, that datasets spanning a much longer time period is required to confirm or disprove this.
 
The surface differential rotation parameter of $\alpha=0.036\pm0.007$ fits well to the observation that rapidly rotating subgiants exhibit relatively low surface shear. \citet{haru_huvir} measured $\alpha=-0.029\pm0.005$ on HU\,Vir, another K subgiant in a binary. On the K subgiant IL\,Hya, \cite{kovari_ilhya} reported $\alpha\approx$0.03. Our results are also in good agreement with the empirical relation of $|\alpha| \approx 0.013P_{\mathrm{rot}}$ for binaries originally suggested by \citet{kovari_diffrot} (see also the right panel of our Fig.~\ref{fig:ccf_eieri}). 

\section{Summary}
Our study of the main compononent of the single lined RS CVn binary EI Eri based on a 4 decades long photometric time series and a multi-site high resolution spectroscopic campaign yielded the following results:
\begin{itemize}
    \item Based on a 41 year long photometric dataset, we derive a roughly 2.5, a 4.5--5.5 and a 8.9--11.6 year photometric cycle. The apparent periodicity in the seasonal changes of the rotational period coincides with the longest cycle. The long term $B-V$ and $V-I$ multicolor photometry indicates spot dominated activity, rather than plage dominated.
    \item The FFD of 41 flares shows a broken power law shape with power law indices of $1.41\pm0.02$ and $2.26\pm0.08$. According to \cite{2018ApJ...854...14M}, this might indicate that there is a critical energy break point on the FFD, where the flaring loop size exceeds the local scale height, resulting in higher energy flares and steeper power law indices. Phase distribution of the flares indicate that magnetic flux emergence is affected by tidal effects \citep{2003A&A...405..291H, holzandschuss2003}, and flares are concentrated on the hemisphere facing the secondary.
    \item Using our MUSICOS spectral time series, we derived four consecutive, independent Doppler-images of EI\,Eri, covering 1.8, 1.6, 1.0 and 2.9 rotations, enabling us to study very short term spot evolution on the rapidly evolving surface. The resolution of the spectra were blurred down to the lowest available resolution in order to avoid artifacts caused by different resolution. This approach was validated by a series of tests. The Doppler images revealed a strong, evolving, but always present polar cap and several emerging and disappearing low latitude features of constantly evolving shape and contrast. The approximate lifetime of these structures coincide with the lower limit proposed by \cite{namekata_spot_lifetime}, although datasets covering longer time periods are needed in order to conclusively verify this.
    \item Cross correlating the consecutive Doppler images revealed a weak, solar-like surface differential rotational pattern of $\alpha=0.036\pm0.007$, which is in good agreement with the empirical law for binaries proposed by \cite{kovari_diffrot}.

\end{itemize}
Our paper also includes tests to show that our multi-instrument time-series can be used for Doppler imaging without introducing artifacts, if treated properly.

\begin{acknowledgements}
The authors would like to thank Albert Washuettl and the original MUSICOS team for organizing and gathering the data used in this work.
The authors acknowledge the Hungarian National Research, Development and Innovation Office
grants OTKA K-131508, KKP-143986 (\'Elvonal), and 2019-2.1.11-T\'eT-2019-00056. LK acknowledges the Hungarian National Research, Development and Innovation Office
grant OTKA PD-134784. LK and KV are Bolyai J\'anos research Fellows.
KV is supported by the Bolyai+ grant \'UNKP-22-5-ELTE-1093, BS is supported by the \'UNKP-22-3 New National Excellence Program of the Ministry for Culture and
Innovation from the source of the National Research, Development and Innovation Fund. GWK acknowledges long-term support from NASA, NSF, and the State of
Tennessee through its Centers of Excellence Program.
\end{acknowledgements}


\bibliography{eieri}
\clearpage
\begin{appendix}
\section{Resolution tests}\label{sect:diffrottest}
\label{sect:test}
\begin{figure*}[b!!!]
\begin{multicols}{2}
    \includegraphics[width=\linewidth]{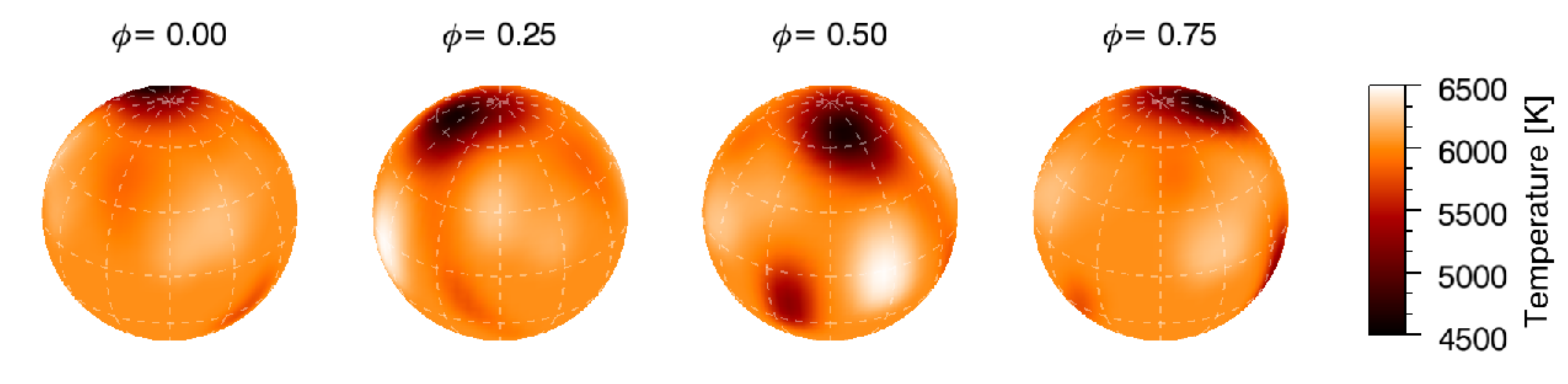}\par 
    \includegraphics[width=\linewidth]{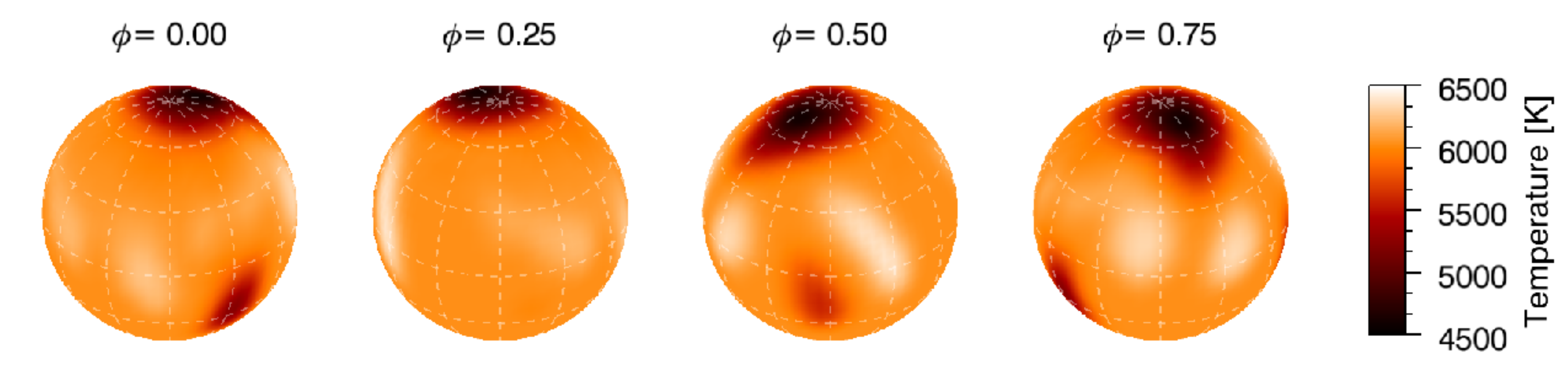}\par 
    \end{multicols}
\begin{multicols}{2}
    \includegraphics[width=\linewidth]{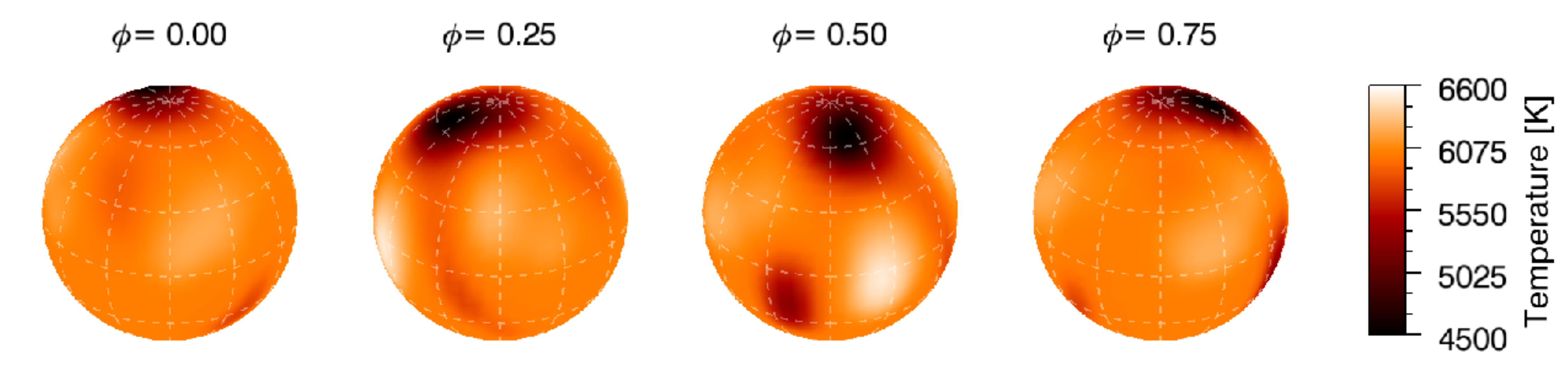}\par 
    \includegraphics[width=\linewidth]{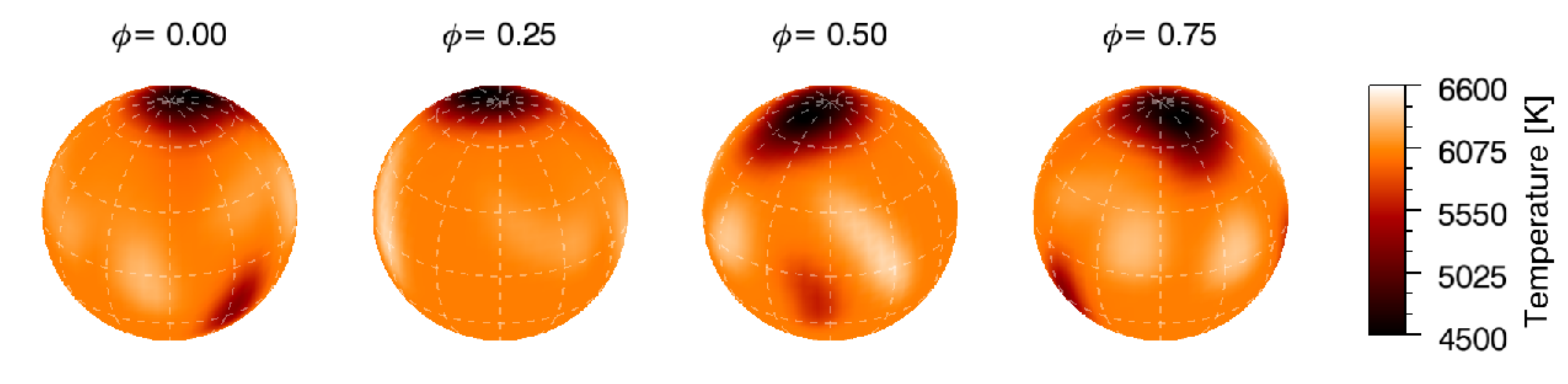}\par 
    \end{multicols}
\begin{multicols}{2}
    \includegraphics[width=\linewidth]{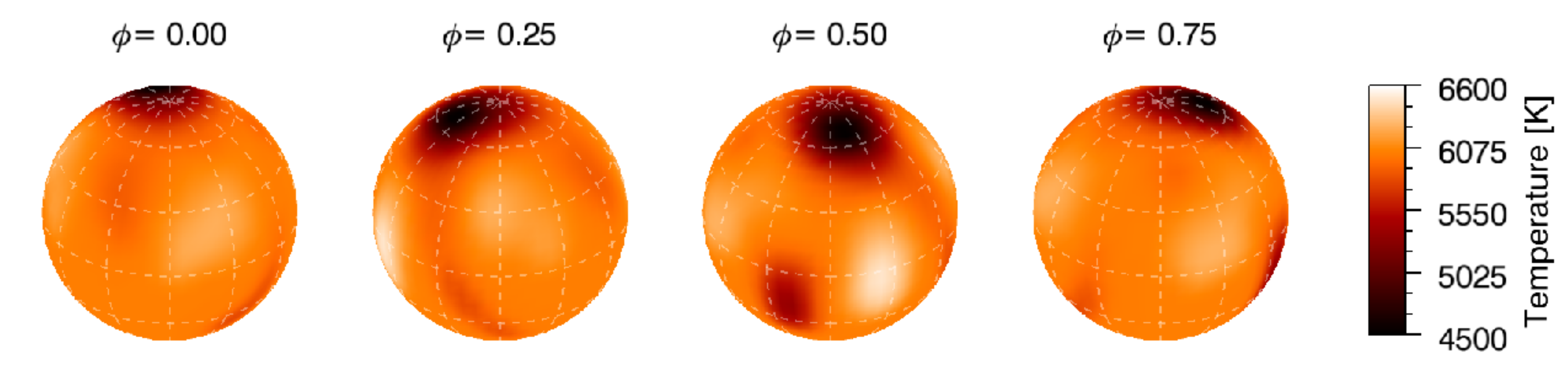}\par
    \includegraphics[width=\linewidth]{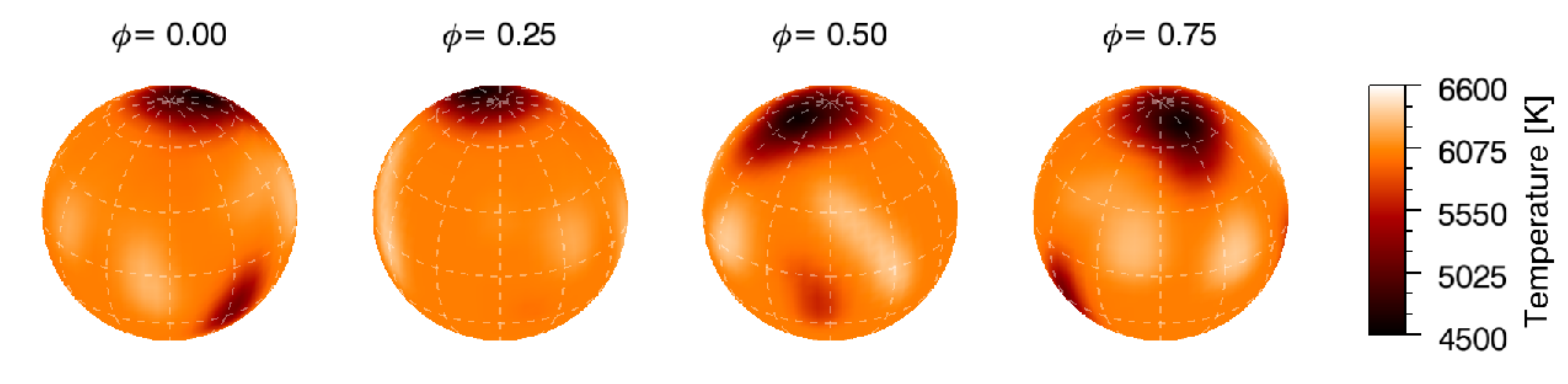}\par
\end{multicols}
\begin{multicols}{2}
    \includegraphics[width=\linewidth]{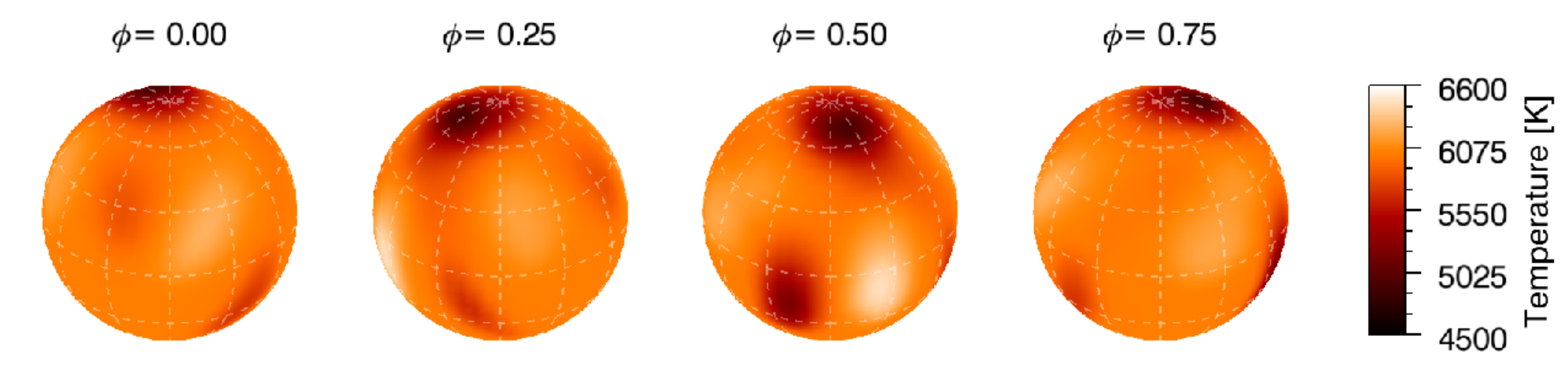}\par
    \includegraphics[width=\linewidth]{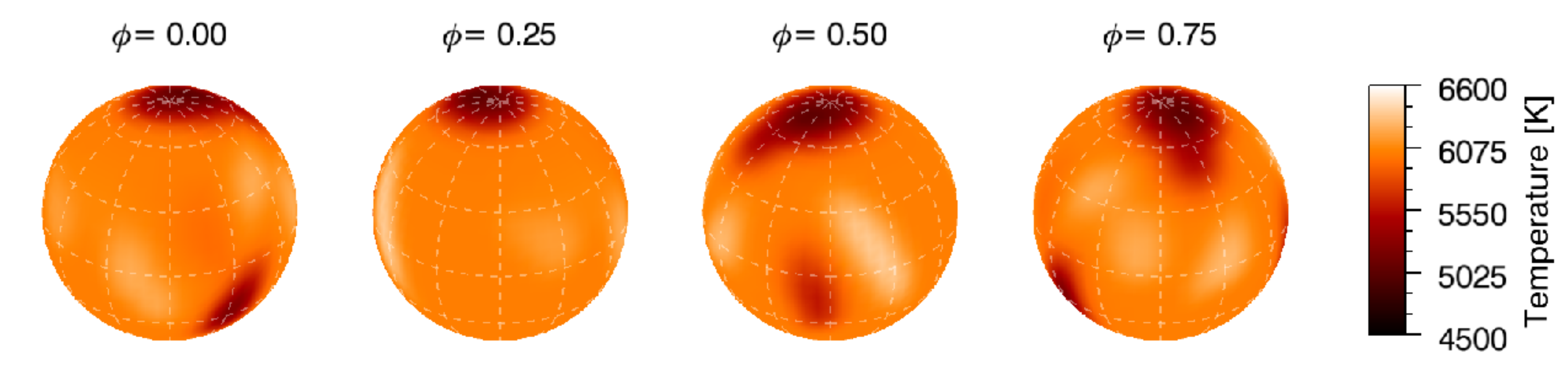}\par
\end{multicols}
\caption{Doppler imaging test for two consecutive rotations of V1358\,Ori. Images on the left correspond to the first rotation, while images on the right correspond to the second one. The four rows from top to bottom correspond to data of different spectral resolutions: $R=80\,000$ \citep[original data taken from][]{kriskovics2019}, and reduced levels of $R=60\,000$, $R=40\,000$ and $R=20\,000$, respectively.}
\label{fig:ditest_maps}
\end{figure*}
\begin{figure*}[b!]
\begin{multicols}{2}
    \includegraphics[width=\linewidth]{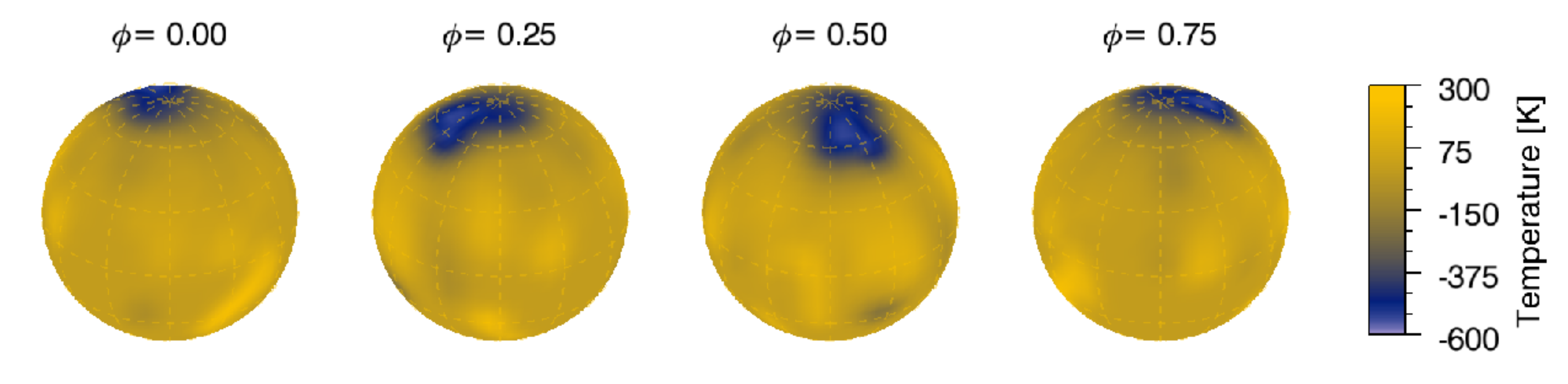}\par 
    \includegraphics[width=\linewidth]{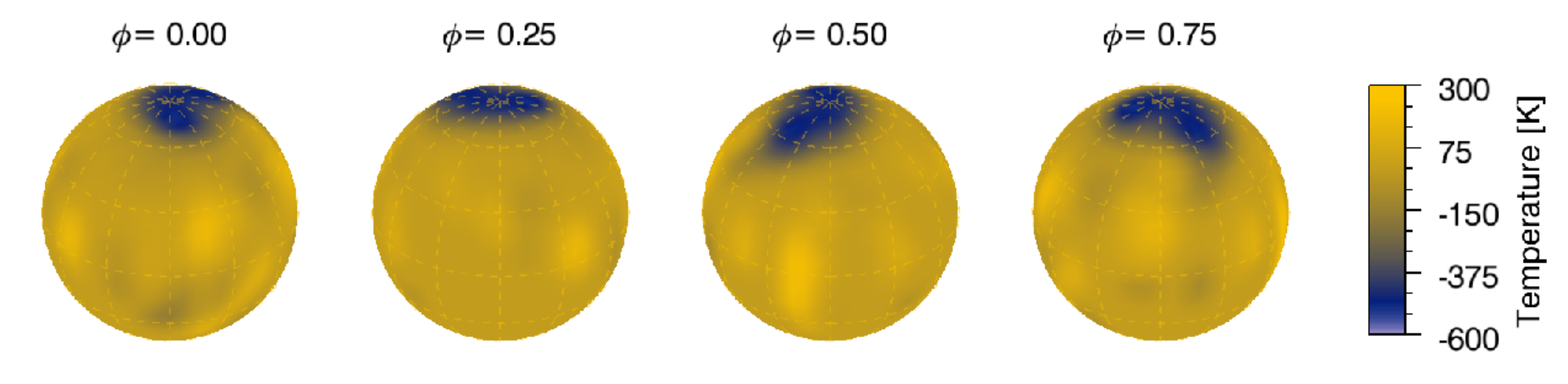}\par 
\end{multicols}
\caption{Difference images of the original V1358\,Ori Doppler images \citep{kriskovics2019} and those that correspond to a spectral resolution reduced to $R$=20\,000. The left panel corresponds to the first rotation and the right panel for the second, cf. Fig.~\ref{fig:ditest_maps}.}
\label{fig:ditest_maps_diff}
\end{figure*}
MUSICOS 1998 data of EI\,Eri were obtained from seven different instruments with different spectral resolutions (and hence, there are spectra with different resolutions in each subset), which can induce artifacts on the Doppler maps due to the different levels of detail of the spectral features and the different instrumental profiles. To counteract this, we  used a Gaussian kernel to decrease the resolution of the spectra to our lowest available resolution. Since this has not been done before, we carry out tests to see the effect of this treatment on the high resolution dataset of V1358\,Ori \citep{kriskovics2019}. We created datasets of decreased resolution corresponding to  $R$=20\,000, 30\,000, 40\,000, 50\,000, 60\,000 and 70\,000, and compared it to the original maps. For the details on the original dataset, astrophysical parameters and inversion, we refer to \cite{kriskovics2019}.

Doppler images for the datasets of decreased resolutions were derived with the same inversion settings and astrophysical parameters. Hotter features were not suppressed on purpose, to see if artifacts are introduced on lower resolutions. Fig.\,\ref{fig:ditest_maps} shows examples for both rotations corresponding to $R$=60\,000, 40\,000 and 20\,000, along with the originals. Visual comparison of the test maps and the original ones reveal that the overall structure did not change on the maps corresponding to lower spectral resolutions, except for a few smaller features that gradually faded as the resolution decreased. Spot temperatures however can significantly change in the case of the the strongest active regions, especially on the maps corresponding to $R$=20\,000. This is expected, since the blurring makes the bumps caused by the spots on the line profiles shallower. Fig.\,\ref{fig:ditest_maps_diff} shows the difference maps of the original and the $R=20\,000$ map. It is apparent that in the case of the strong polar feature, there is about 600\,K difference in the recovered spot temperature, while smaller features did not change much. Finally, it is important to note that artifacts are not introduced by the lower resolution.

Based on these tests we conclude that degrading the MUSICOS spectra to a reduced spectral resolution of $R$=20\,000 is still sufficient to investigate the rapid spot evolution and differential rotation of EI\,Eri, despite the fact that there is some fading of the spectral features due to a decrease in contrast, which affects the measurement of the differential rotation at some level.

\begin{figure*}[t!!!]
\begin{multicols}{2}
    \includegraphics[width=\linewidth]{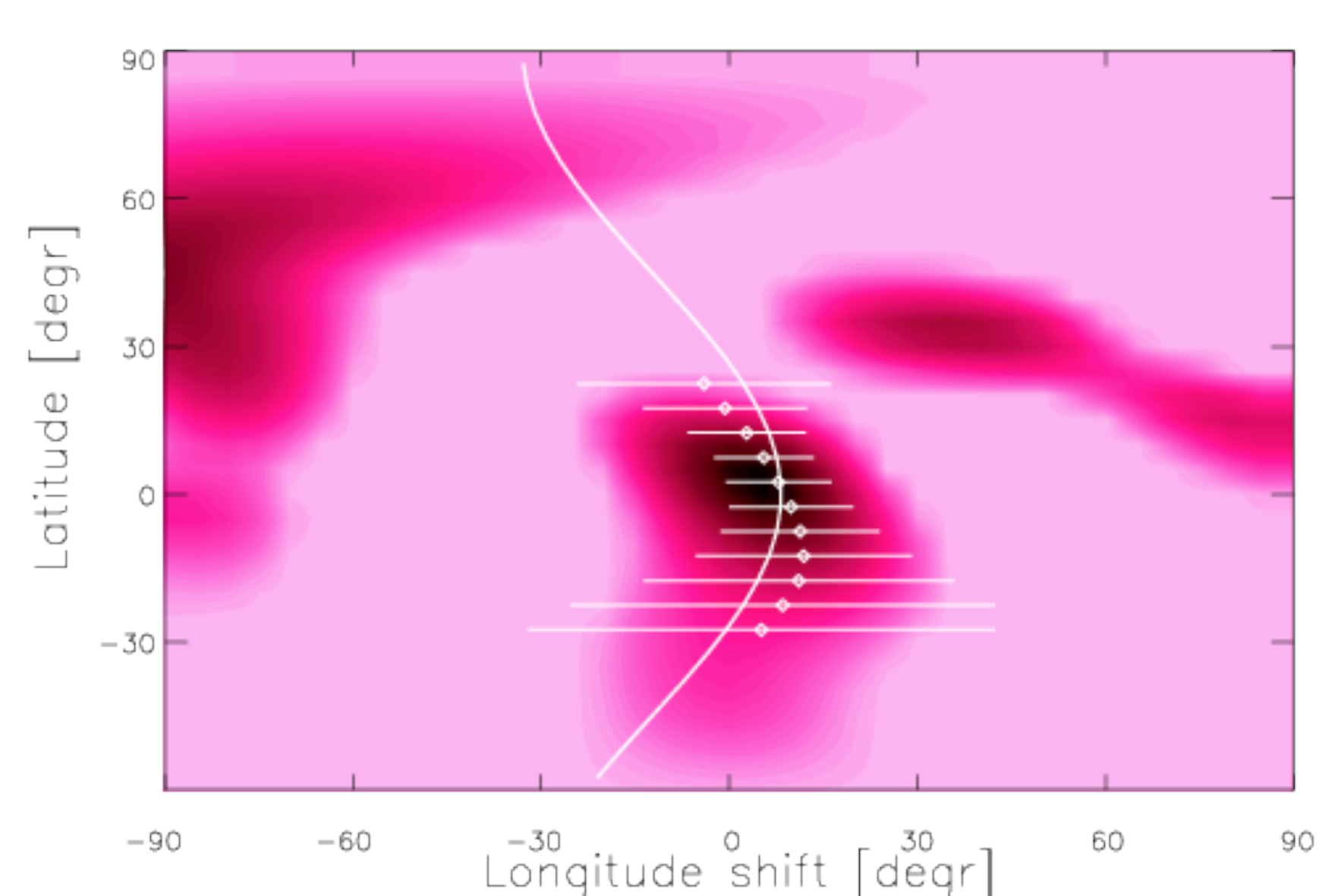}\par 
    \includegraphics[width=\linewidth]{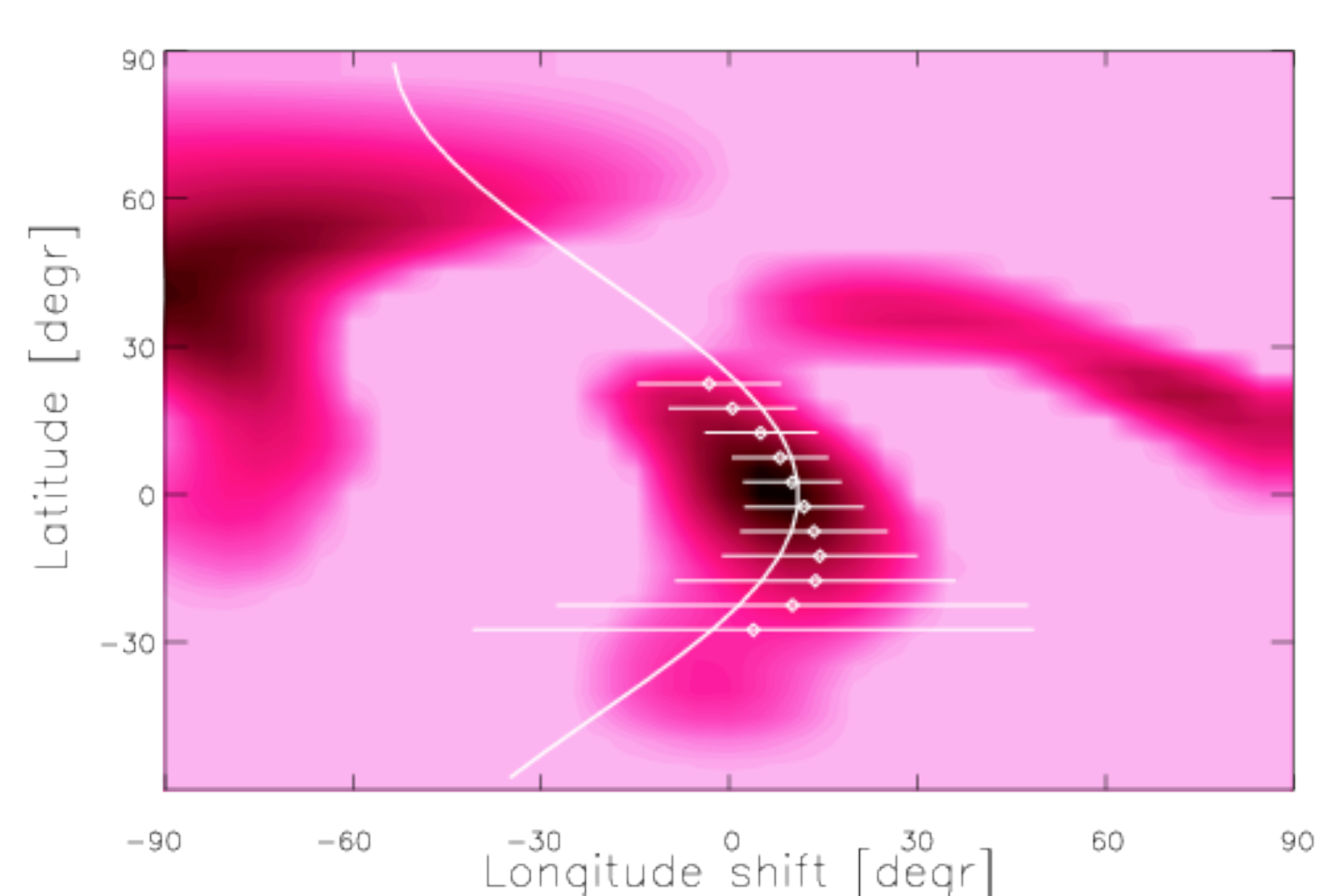}\par 
\end{multicols}
\caption{Example cross-correlation test maps of V1358\,Ori with their fitted surface differential rotation functions obtained for different spectral resolutions of $R$=20\,000 (left) and $R$=70\,000 (right). The derived surface shear coefficients are $\alpha=0.010\pm0.0025$ and $\alpha=0.015\pm0.003$, respectively. For the original cross-correlation map see \citet{kriskovics2019}.}
\label{fig:drtest}
\end{figure*}

\begin{figure}[t!!!]
    \includegraphics[width=\linewidth]{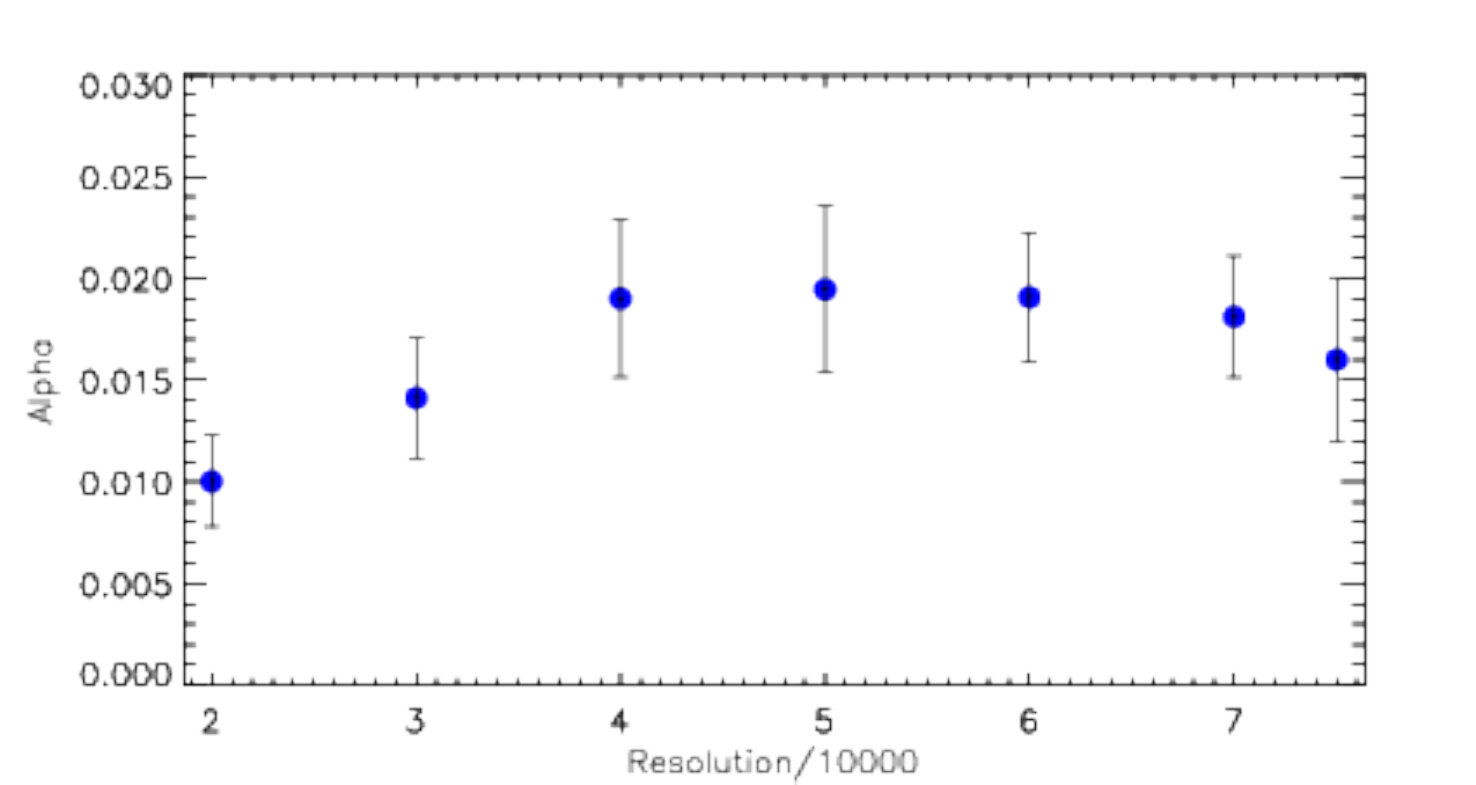}\par 
\caption{The $\alpha$ surface differential rotation parameter values obtained for spectroscopic test datasets with different levels of $R$ spectral resolution.}
\label{fig:drtest_curve}
\end{figure}

The reliability of measuring differential rotation by cross-correlation of consecutive Doppler maps is based on the reliability of the spot reconstruction (for the details of the cross-correlation technique, see Sect.~\ref{sect:diffrot}). Therefore, here we test how the reduced spectral resolution affects the surface shear parameter derived from cross-correlation. For the test the two consecutive rotations of V1358\,Ori (see Fig.~\ref{fig:ditest_maps}) were cross-correlated with different levels of spectral resolution. Two example cross-correlation maps are shown in Fig.~\ref{fig:drtest} which correspond to $R$=20\,000 ($\alpha=0.01 \pm 0.0025$) and $R$=70\,000 ($\alpha=0.015\pm 0.003$). We note that the original cross-correlation map presented in \cite{kriskovics2019} with a spectral resolution of $R$=80\,000 resulted in $\alpha$=0.016. The derived $\alpha$ surface shear values are plotted against the spectral resolution in Fig.~\ref{fig:drtest_curve}. It is apparent that the resolution affects the derived shear coefficient in a unique way, that is, at low resolution the dependence has some linear trend, while at mid to high resolution ($R$>40\,000) there is no real trend, the derived $\alpha$ values agree with each other within the estimated error bars. Nevertheless, on our lowest resolution ($R$=20\,000), the measured surface differential rotation underestimates the actual value by $\approx$60\%.

\clearpage

\onecolumn
\section{Observing log}
\begin{table}[h!!!!]
\begin{threeparttable}
\caption{Observing log of the MUSICOS 1998 data. }
\begin{tabular}{cccc|cccc}
\noalign{\smallskip}
\hline
\noalign{\smallskip}
$HJD$ [d] & Telescope & Rotational phase & Used & $HJD$ [d] & Telescope & Rotational phase & Used \\
\noalign{\smallskip}
\hline
\noalign{\smallskip}
2451141.3758 & OHP & 0.157 & yes &  2451148.7618 & ESO152 & 0.950 & yes \\
2451141.4856 & OHP & 0.213 & yes &  2451148.9391 & MSO & 0.041 & yes \\
2451141.6050 & OHP & 0.274 & yes &  2451149.0746 & MSO & 0.111 & no \\
2451141.6302 & KPNO & 0.287 & yes & 2451149.1508 & MSO & 0.150 & yes \\
2451141.6797 & KPNO & 0.313 & yes & 2451149.2564 & BXO & 0.204 & no \\
2451141.7891 & KPNO & 0.369 & no  & 2451149.5683 & ESO90 & 0.364 & yes \\
2451141.9259 & KPNO & 0.439 & yes & 2451149.7026 & ESO152 & 0.433 & yes \\
2451142.6193 & KPNO & 0.795 & yes & 2451149.7849 & ESO152 & 0.475 & yes \\
2451142.7057 & KPNO & 0.840 & yes & 2451150.1065 & BXO & 0.640 & no \\
2451142.8025 & KPNO & 0.889 & no  & 2451150.2155 & BXO & 0.696 & yes \\
2451142.9096 & KPNO & 0.944 & yes & 2451150.2784 & BXO & 0.729 & yes \\
2451142.9849 & KPNO & 0.983 & no  & 2451150.5648 & ESO90 & 0.876 & yes \\
2451143.7458 & ESO152 & 0.374 & no & 2451150.6669 & ESO152 & 0.928 & yes \\
2451143.8030 & KPNO & 0.403 & yes & 2451150.7474 & ESO152 & 0.970 & yes \\
2451143.8529 & ESO152 & 0.429 & no & 2451150.9827 & MSO & 0.090 & yes \\
2451143.8823 & KPNO & 0.444 & no & 2451151.0459 & MSO & 0.123 & yes \\
2451143.8973 & KPNO & 0.452 & yes & 2451151.1061 & MSO & 0.154 & no \\
2451143.9122 & KPNO & 0.459 & yes & 2451151.5633 & ESO90 & 0.389 & yes \\
2451144.3730 & OHP & 0.696 & yes & 2451151.7115 & ESO152 & 0.465 & yes \\
2451144.5975 & ESO152 & 0.811 & no & 2451151.7969 & ESO152 & 0.509 & yes \\
2451144.7046 & KPNO & 0.866 & yes & 2451152.0938 & BXO & 0.661 & yes \\
2451144.7232 & KPNO & 0.876 & yes & 2451152.1989 & BXO & 0.715 & yes \\
2451144.7419 & KPNO & 0.885 & yes & 2451152.2799 & BXO & 0.757 & yes \\
2451144.8528 & ESO152 & 0.942 & no & 2451152.5011 & INT & 0.870 & no \\
2451144.8673 & KPNO & 0.950 & yes & 2451152.5616 & ESO90 & 0.901 & yes \\
2451144.8859 & KPNO & 0.959 & yes & 2451153.5985 & ESO90 & 0.434 & no \\
2451144.9045 & KPNO & 0.969 & yes & 2451154.5383 & INT & 0.916 & no \\
2451145.3744 & OHP & 0.210 & yes  & 2451154.5934 & ESO90 & 0.945 & yes \\
2451145.6007 & ESO152 & 0.326 & yes & 2451155.1511 & BXO & 0.231 & no \\
2451145.8042 & ESO152 & 0.431 & yes & 2451155.2465 & BXO & 0.280 & no \\
2451146.3699 & OHP & 0.722 & yes & 2451155.3879 & INT & 0.353 & yes \\ 
2451146.5807 & ESO90 & 0.830 & yes & 2451155.5990 & ESO90 & 0.461 & yes \\
2451146.7466 & ESO152 & 0.915 & yes & 2451155.6447 & INT & 0.485 & no \\
2451146.8321 & ESO152 & 0.959 & yes & 2451156.3677 & INT & 0.856 & no \\
2451147.0404 & MSO & 0.066 & yes & 2451156.4359 & INT & 0.891 & yes \\
2451147.0592 & MSO & 0.076 & no  & 2451156.5991 & ESO90 & 0.975 & yes \\
2451147.1641 & MSO & 0.129 & yes & 2451156.6333 & INT & 0.992 & no \\
2451147.4508 & OHP & 0.277 & yes & 2451157.3997 & INT & 0.386 & no \\
2451147.5618 & ESO90 & 0.334 & yes & 2451157.4703 & INT & 0.422 & no \\
2451147.7280 & ESO152 & 0.419 & yes & 2451157.5991 & ESO90 & 0.488 & no \\
2451147.8123 & ESO152 & 0.462 & yes & 2451157.6475 & INT & 0.513 & no \\
2451147.9404 & MSO & 0.528 & no & 2451158.5358 & INT & 0.969 & no \\
2451148.0240 & MSO & 0.571 & yes & 2451158.5984 & ESO90 & 0.001 & no \\
2451148.1543 & MSO & 0.638 & no & 2451159.5998 & ESO90 & 0.516 & no \\
2451148.6718 & ESO152 & 0.904 & yes & 2451160.6313 & ESO90 & 0.045 & no \\
\noalign{\smallskip}
\hline
\end{tabular}
\label{tab:obslog}
\begin{tablenotes}
    \item \small Columns from left to right: the Heliocentric Julian Dates ($HJD$), the observing facilities, the rotational phases (according to Eq.~\ref{eq1}), and whether the measurement was used or not.
\end{tablenotes}
\end{threeparttable}
\end{table}

\clearpage
\onecolumn
\section{Line profile fits}
\begin{figure*}[h!!!!]
    \begin{multicols}{4}
    
    \includegraphics[width=\linewidth]{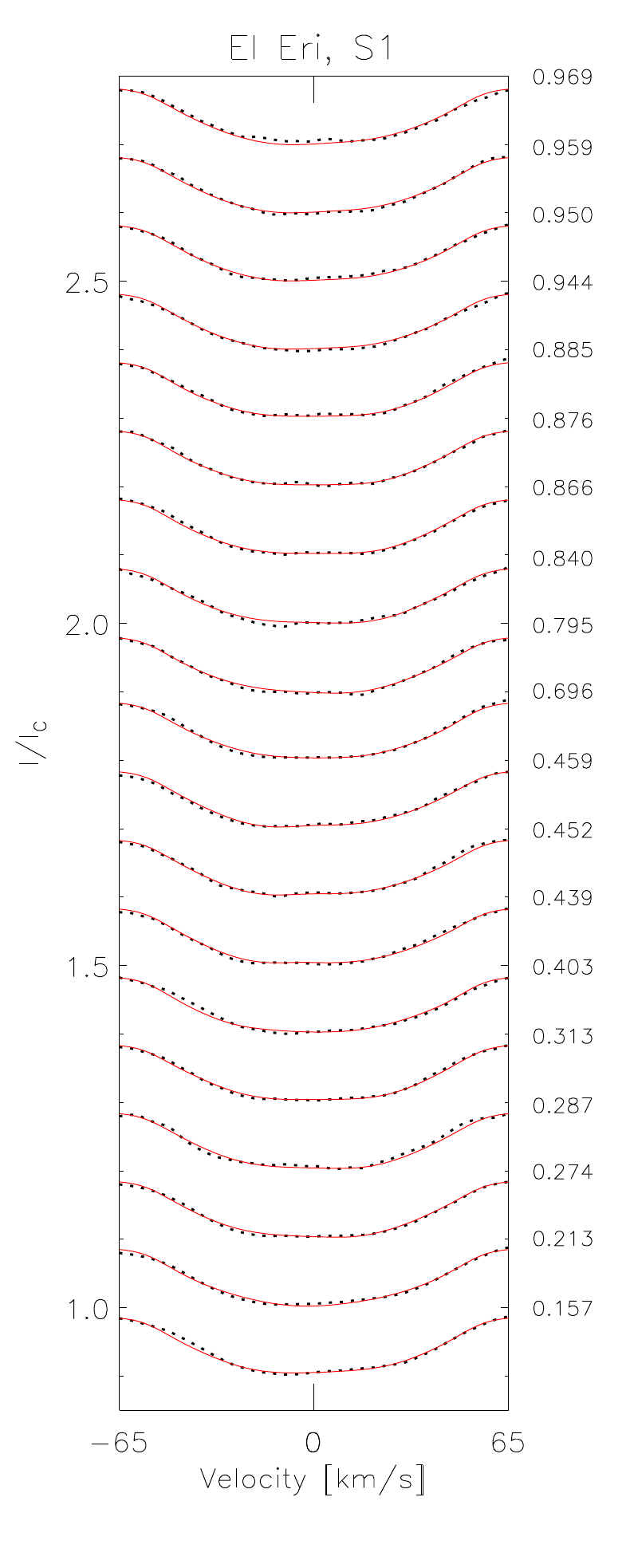}\par 
    \includegraphics[width=\linewidth]{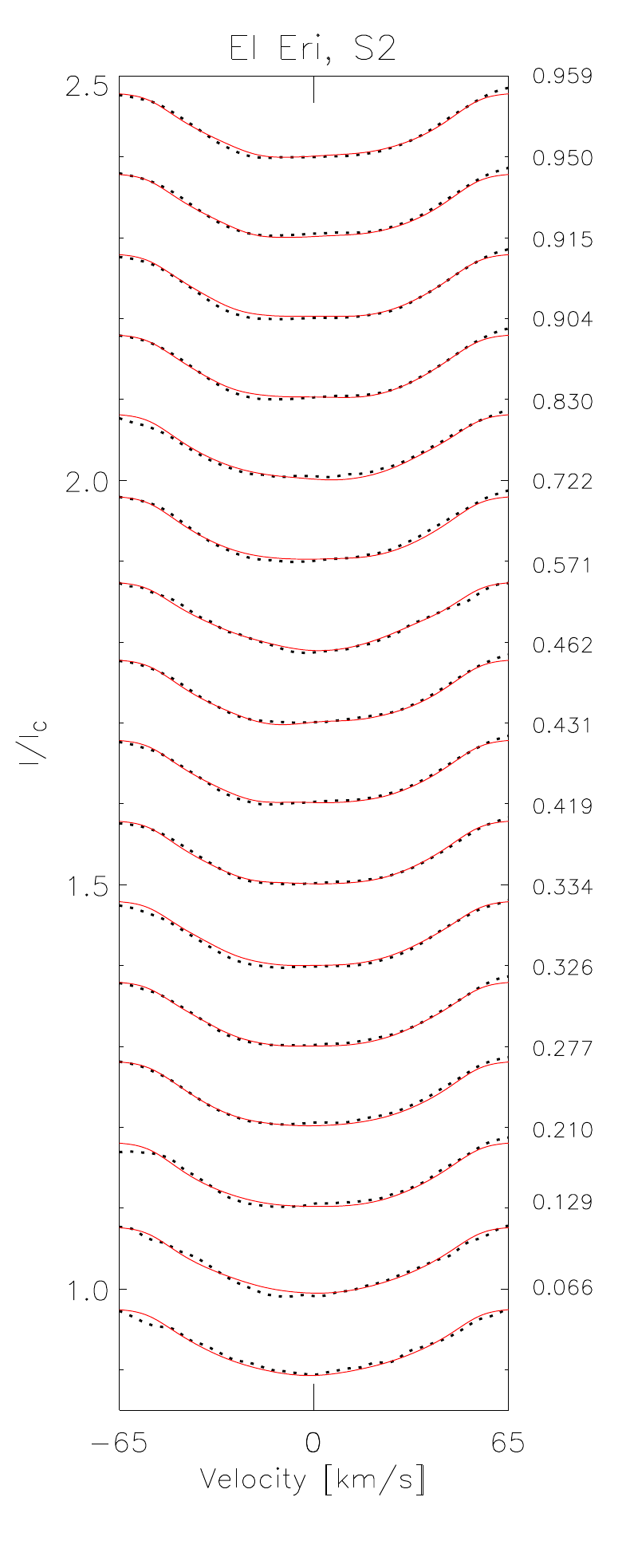}\par
    \includegraphics[width=\linewidth]{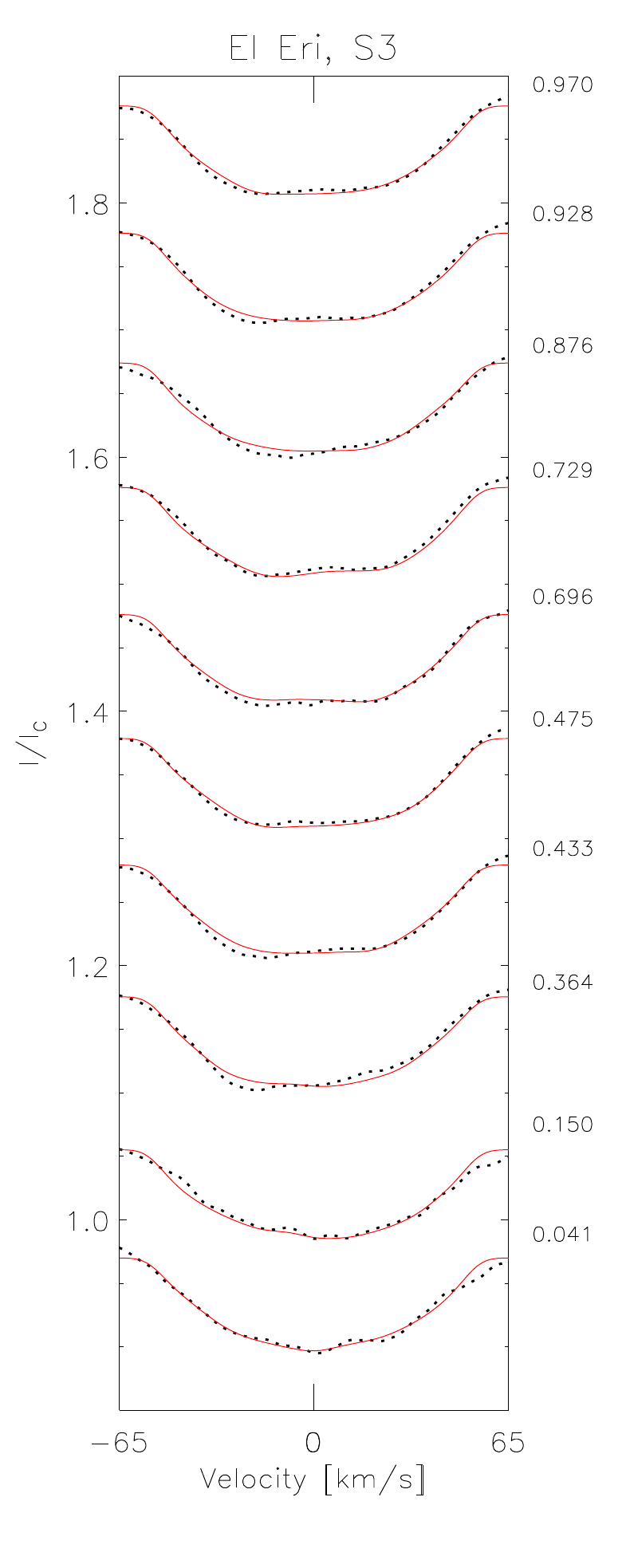}\par 
    \includegraphics[width=\linewidth]{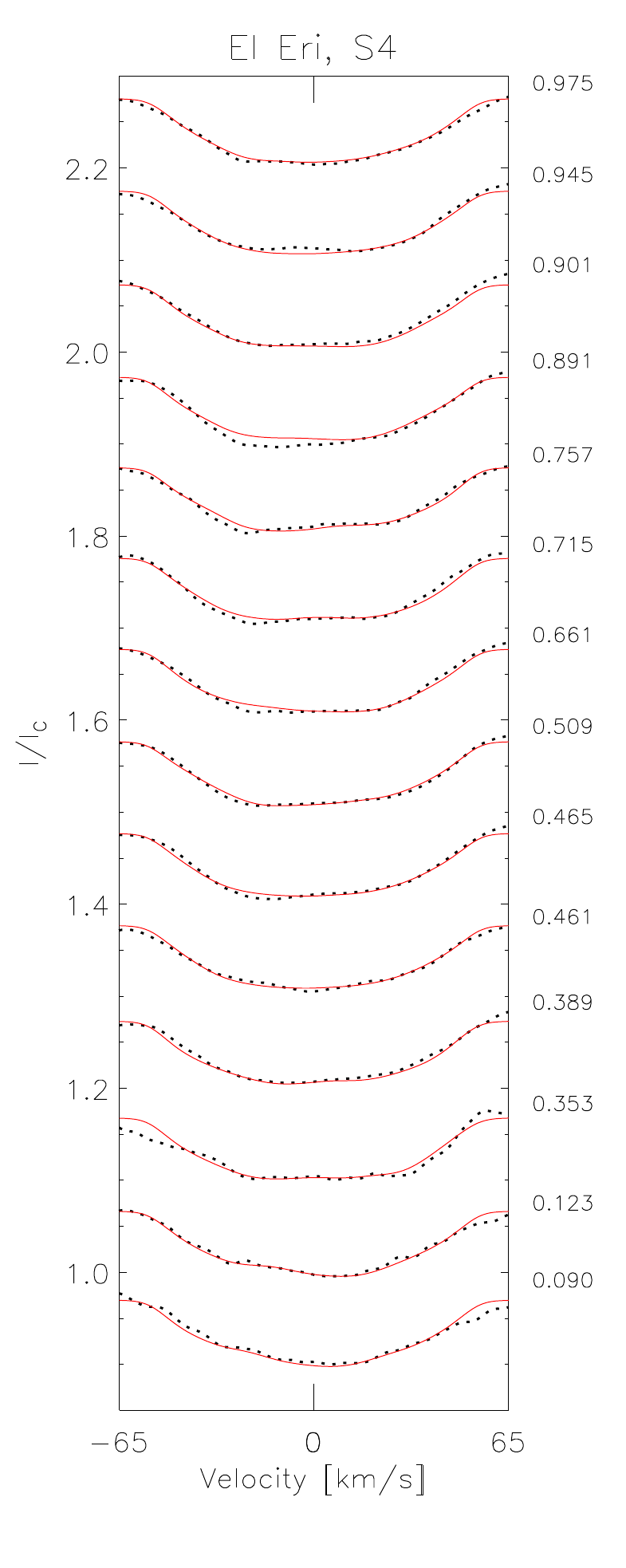}\par
    
    \end{multicols}
\caption{Fitted line profile subsets (S1-S4) corresponding to the four Doppler images (DI1-DI4) presented in Fig.~\ref{fig:di_eieri}. The phase distributions of each subset are shown in the right panel of Fig.~\ref{fig:obsplots}.}
\label{fig:di_eieri_profs}
\end{figure*}
\end{appendix}

\end{document}